\documentclass[aps,twocolumn,superscriptaddress,nofootinbib,a4paper,longbibliography]{revtex4-1}

\usepackage{times}
\usepackage{epsfig}
\usepackage{amsfonts}
\usepackage{amsmath}
\usepackage{esint}
\usepackage{amsthm}
\usepackage{amssymb}					
\usepackage{amsthm}
\usepackage{dsfont}
\usepackage{bm}
\usepackage{mathtools}
\usepackage{color}
\usepackage{multirow}
\usepackage[normalem]{ulem}
\newcommand{\stkout}[1]{\ifmmode\text{\sout{\ensuremath{#1}}}\else\sout{#1}\fi}
\usepackage{latexsym}
\usepackage{mathrsfs}
\usepackage{natbib}
\usepackage{verbatim}
\usepackage[T1]{fontenc}
\usepackage{float}
\usepackage{graphicx}
\usepackage{subcaption}
\usepackage{xcolor}
\usepackage{soul}
\usepackage{caption}
\usepackage{subcaption}
\usepackage{enumerate}
\usepackage[font=small,labelfont=bf]{caption}

\usepackage{microtype}

\usepackage{xspace}

\newtheorem{result}{Result} 

\newcommand{\elna}[1]{\textit{\small\textcolor{teal}{Elna: #1}}}

\theoremstyle{definition}

\DeclareMathOperator{\tr}{tr}
\newcommand{\ket}[1]{|#1\rangle}
\newcommand{\bra}[1]{\langle#1|}
\newcommand{\bracket}[3]{\langle#1|#2|#3\rangle}

\usepackage{amssymb}
\usepackage{amsthm}
\usepackage{mathtools} 

\usepackage[colorlinks=true,linkcolor=blue,citecolor=magenta,urlcolor=blue]{hyperref}

\begin{document}
	

\title{Quantum inputs in the prepare-and-measure scenario and stochastic teleportation}

\author{Elna Svegborn}\thanks{elna.svegborn@fysik.lu.se}
\affiliation{Physics Department and NanoLund, Lund University, Box 118, 22100 Lund, Sweden.}

\author{Jef Pauwels}
\affiliation{Department of Applied Physics, University of Geneva, Switzerland.}
\affiliation{Constructor Institute of Technology (CIT), Switzerland.}
\affiliation{Constructor University, 28759 Bremen, Germany.}

\author{Armin Tavakoli}
\affiliation{Physics Department and NanoLund, Lund University, Box 118, 22100 Lund, Sweden.}

\begin{abstract}
We investigate prepare-and-measure scenarios in which a sender and a receiver use entanglement to send quantum information over a channel with limited capacity. We formalise this framework, identify its basic properties and provide numerical tools for optimising quantum protocols for generic communication tasks. The seminal protocol for sending quantum information over a classical channel is teleportation. We study a natural stochastic generalisation in which the sender holds $N$ qubits from which the receiver can recover one on demand. We show that, if the classical communication is allowed to exploit extremal non-signaling correlations, then two bits of communication suffice to solve this task exactly for any $N$. We then consider entanglement-based protocols and show that these can be constructed systematically by leveraging connections to several well-known quantum information primitives, such as teleportation, cloning machines and random access coding. In particular, we show that by using genuine multi-particle entangled measurements, one can construct a universal stochastic teleportation machine, i.e.~a device whose teleportation fidelity is independent of the quantum input. 
\end{abstract}

\date{\today}
\maketitle

\section{Introduction}
The transfer of information from one party to another requires communication. The most common stage for investigating the physics of this process is the prepare-and-measure (PM) scenario. In the PM scenario, a sender holds some information, encodes it into a message of limited alphabet, and sends it to the receiver. The receiver then decodes the message  to extract a particular property of the input, \emph{a priori} unknown to the sender. Today, it is well established that quantum resources can both enable and enhance communication beyond classical limits.


Quantum resources can be introduced in the PM scenario in several inequivalent ways. The most straightforward way is to upgrade the message from being a classical symbol to a quantum state. Decoding the quantum message via a quantum measurement leads to higher-than-classical rates of successful communication. This has been showcased in  well-known tasks such as random access codes \cite{Ambainis1999, Nayak1999, Tavakoli2015} and quantum dimension witnesses \cite{Brunner2013, Aguilar2018, Giordani2023, Bernal2024, Hakanson2025}. A second option is to send only classical messages but let the sender and receiver share quantum entanglement. Entanglement-assisted classical communication is fundamentally propelled by quantum nonlocality \cite{Brukner2004, Tavakoli2020, Pauwels2022}, and its advantages can be either larger \cite{Pawlowski2010} or smaller \cite{Martinez2018} than using quantum messages. A powerful synergy is to combine both quantum resources, i.e.~to send quantum messages assisted by shared entanglement \cite{Tavakoli2021, Pauwels2022b}. The seminal example of this is dense coding, in which the classical capacity of the channel is doubled \cite{Bennett1992}, and recent works have found that significant advantages can be harvested already with single-particle quantum measurements \cite{Piveteau2022, Bakhshinezhad2024, Zhang2025}.

However, it is crucial to distinguish whether the information being transmitted is classical or quantum—that is, whether the sender's input consists of classical symbols or quantum states. While the above discussion focuses on the former,
 \begin{figure}[ht!]
	\centering
	\includegraphics[width=\columnwidth]{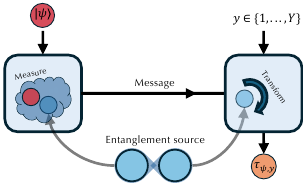}
	\caption{\textit{Entanglement-assisted prepare-and-measure scenario with quantum inputs.} Alice selects quantum input data $|\psi\rangle$ from a set $\mathcal{I}$ which she encodes in an entanglement-assisted message sent to Bob. Upon receiving the message, Bob selects a classical input $y$, performs a transformation on his particles and outputs the quantum state $\tau_{\psi,y}$.}\label{Fig_scenario}
\end{figure}
it is also important to consider the latter. 
The most well-known example of a PM scenario with quantum inputs is teleportation \cite{Bennett1993}: it allows the sender to send one qubit of information to the receiver by using shared entanglement and two bits of classical communication. While the task of teleportation has been intensely studied \cite{Pirandola2015, Hu2023}, little is known about the quantum communication capabilities of physical systems beyond the specific case of teleportation.


In this paper we adress this question by analysing PM scenarios with quantum inputs. We begin by formalising the entanglement-assisted PM scenario with quantum inputs for the sender; see Fig~\ref{Fig_scenario}. There are two ubiquitous but distinct types of communication: when the message is classical and when it is quantum. We show that the two have a simple and general relation. Specifically, if the parties have access to unlimited entanglement, the capacity of quantum communication becomes equivalent to that of sending twice the amount of classical messages. This motivates us to focus on the practically simpler case of classical communication. In order to quantify the performance of generic protocols, we discuss relevant benchmarks and show how they can be systematically optimised over the available quantum resources by means of tailored alternating convex search algorithms.

We then apply this framework to explore quantum communication tasks involving quantum inputs. We focus on a natural generalization of teleportation, which we term stochastic teleportation.  In stochastic teleportation, the sender holds $N$ separate $d$-level systems (qudits) while the receiver randomly selects a classical input $y\in\{1,\ldots,N\}$, corresponding to the qudit he wishes to learn; see e.g.~Refs~\cite{Pitalua2013, Sakharwade2023}. The sender does not know $y$, which is privately selected by the receiver, and must therefore try to generate a quantum state in the receiver's lab from which \textit{any} one of the $N$ qudits can be recovered. To make this possible, the parties are allowed the same resources as in standard teleportation ($N=1$), namely shared entanglement and two dits of classical communication.  The stochastic teleportation task is distinct from a multi-particle teleportation because only partial information about the system is teleported, and the choice of which partial information to teleport is made remotely. The task can alternatively be viewed as a quantum random access code, but with the key distinction that the information being randomly accessed is itself a quantum state. This task serves as a natural setting for studying quantum data compression and the communication advantages enabled by quantum information.

It was shown in Ref.~\cite{Grudka2015} that the simplest stochastic teleportation scenario, namely for a pair of qubits, can be performed exactly if the parties are granted access to a Popescu-Rohrlich nonlocal box for communication \cite{Popescu1994}.We show that two dits of classical communication, augmented by extremal no-signaling correlations, are sufficient to implement stochastic teleportation exactly for all dimensions $d$ and all $N$, effectively trivialising the task. This can be viewed as a quantum information manifestation of the exceptional capabilities of post-quantum nonlocality \cite{vanDam2013}. A natural next step is to study quantum protocols for stochastic teleportation. We show a systematic way to construct stochastic teleportation protocols by combining standard teleportation and entanglement-assisted random access codes \cite{Pawlowski2010}. For these protocols, we present both simple and general bounds as well as an optimal solution for the simplest stochastic teleportation scenario. However, their entanglement cost is significant, because they require entanglement to be consumed first for teleportation and then for entanglement-assisted classical communication.


Finally, we take a conceptually different route by asking whether there exists universal protocols for stochastic teleportation, i.e.~protocols whose performance is independent of the quantum input. We show that such protocols are possible and that they require only modest entanglement consumption compared to the above mentioned protocols. For stochastic teleportation of a pair of qubits, we show that a universal teleportation fidelity of $5/6$ is achievable by sharing only one entangled bit. This protocol bears no apparent resemblance to standard teleportation because it is based on performing entangled measurements jointly over three qubits. We also use our numerical tools to study universal protocols for more than two qubits and find not only that these exist but also that their performance can be enhanced by sharing multiple entangled bits between the parties.

\section{Scenario and elementary results}

Consider a PM scenario featuring a sender and a receiver, whom we call Alice and Bob respectively; see Fig~\ref{Fig_scenario}. Alice holds a private quantum input $\ket{\psi}\in\mathcal{H}_{A'}$  selected from a set of states denoted $\mathcal{I}$. This set is part of the description of the scenario and may be selected to either contain finitely many states or infinitaley many states (for example all pure qubit states). Bob privately selects a classical input $y\in\{1,\ldots,Y\}$, based on which he performs an operation that generates a quantum output state $\tau_{\psi,y}$ which belongs to a Hilbert space $\mathcal{H}_{B'}$. To enable this, the parties may share an entangled state $\rho_{AB} \in \mathcal{H}_A \otimes \mathcal{H}_B$ and Alice is allowed to communicate a single message to Bob. This communication can be either a classical or a quantum message with dimension $d_C$. Thus, the scenario is completely specified by the tuple  $\{\mathcal{I},Y,\mathcal{H}_{B'}\}$ which defines the input and output spaces of the parties. Similarly, the available resources are described by the tuple $\{d_C,\rho,R\}$, where $R$ indicates whether the communication is a classical ($R=\mathcal{C}$) or a quantum ($R=\mathcal{Q}$) message. We now consider these two cases separately.

\textbf{Classical communication.} If the communication is classical, Alice's most general strategy consist in jointly measuring her share of $\rho_{AB}$ and her quantum input $\psi_{A'}$. This measurement is represented by a positive operator-valued measure (POVM) denoted $\{M^{c}\}_{c=1}^{d_C}$. The outcome $c$ is sent to Bob who uses it together with his private input $y$ to select a decoding channel $\Lambda_{c,y}:\mathcal{H}_{B}\rightarrow \mathcal{H}_{B'}$ (a completely positive trace-preserving (CPTP) map) which is applied to his half of $\rho_{AB}$. Thus, the final output state becomes
\begin{equation}\label{Ccorr}
	\tau_{\psi,y}= \sum_{c=1}^{d_C} \Lambda^{B\rightarrow B'}_{c,y}\big[\sigma_{c|\psi}\big],
\end{equation}
where $\sigma_{c|\psi}$ is the sub-normalised state remotely prepared by Alice for Bob,
\begin{equation}\label{Cstates}
\sigma_{c|\psi}= \tr_{A'A}\left[(\psi_{A'}\otimes \rho_{AB})(M^c_{A'A}\otimes\openone_B)\right].
\end{equation}

\textbf{Quantum communication.} If the communication is quantum, Alice encodes her message using a quantum channel $\Gamma:\mathcal{H}_{A'}\otimes \mathcal{H}_{A}\rightarrow \mathcal{H}_{C}$ that acts jointly on $\psi_{A'}$ and her half of the entangled state $\rho_{AB}$ and transforms these into a quantum message of Hilbert space dimension $d_C=\text{dim}(\mathcal{H}_C)$. Bob uses his private input $y$ to select a decoding channel $\Lambda_y:\mathcal{H}_{B}\otimes\mathcal{H}_C\rightarrow \mathcal{H}_{B'}$. This is a CPTP map which transforms his half of $\rho_{AB}$ and the incoming quantum message into the final quantum output, which reads
\begin{equation}\label{Qcorr}
	\tau_{\psi,y}=\Lambda_y^{BC\rightarrow B'}\left[(\Gamma^{A'A\rightarrow C}\otimes \openone_B)\left[\psi_{A'}\otimes \rho_{AB}\right]\right] \,.
\end{equation}
We refer to the set of states $\{\tau_{\psi,y}\}_{\psi,y}$ as the quantum correlations. This naming convention mirrors the standard terminology of PM scenarios with classical inputs, but we emphasise that in contrast to such scenarios, where the correlations refer to conditional probability distributions, in our scenario the correlations are quantum output states.  Note that the case of quantum messages is strictly more general than the case of classical messages, since the latter can be obtained from the former by selecting the encoding channel $\Gamma^{A'A\rightarrow C}$ to have a classical output.

We note that PM scenarios with classical inputs emerge as special cases of the above more general formalism. For instance, if we select the set $\mathcal{I}$ to consist only of distinguishable states,  the above reduces to the entanglement-assisted PM scenarios introduced in Ref.~\cite{Tavakoli2021}. To see this, we need only to first  measure $\psi_{A'}$ in the basis that distinguishes the elements of $\mathcal{I}$ and then use the outcome as a classical input on the remaining part of the encoding procedure pertaining to system $A$. Similarly, if we also substitute the entangled state $\rho$ for a separable state, the scenario further reduces to the basic PM scenario with classical inputs and shared randomness, which is the focus of most previous literature.

\subsection{Equivalence between quantum and classical communication}
Consider that the parties are allowed to use entanglement freely but that their communication is restricted. What is the relation between the set of correlations that can be generated with classical messages \eqref{Ccorr} and the set of correlations that can be generated with quantum messages \eqref{Qcorr}? We now show that these sets are equivalent if the amount of classical communication is twice the amount of quantum communication.
\begin{result}[Classical vs quantum messages with unbounded entanglement]\label{Result1}
Consider any  scenario in which any amount of entanglement can be used. Any set of correlations $\{\tau_{\psi,y}\}$ is realisable with quantum messages of dimension $d_C$ if and only if it is realisable with classical messages of dimension $d_C^2$.
\end{result}
\begin{proof}
Assume that  $\{\tau_{\psi,y}\}_{\psi,y}$ is realisable with the resources $\{d_C,\rho,\mathcal{Q}\}$. Consider now a protocol with classical communication in which the encoding operation $\Lambda^{A'A\rightarrow C}$ is performed but the system $C$ is kept in the Alice's lab. Instead of relaying it to Bob, the parties use an auxiliary maximally entangled state $\ket{\phi^+_{d_C}}=\sum_{i=0}^{d_C-1}\ket{ii}$ to teleport $C$ into the lab of Bob. This requires  Alice to perform  a $d_C$-dimensional Bell state measurement on her half of $\phi^+_{d_C}$ and system $C$, and then relay the classical outcome, which has a $d_C^2$-sized alphabet. After performing the correction unitary, Bob can implement the same decoding channel $\Gamma_y$ as in the original protocol.

Conversely, assume that  $\{\tau_{\psi,y}\}_{\psi,y}$ is realisable with the resources  $\{d^2_C,\rho,\mathcal{C}\}$. Consider now a protocol with quantum communication in which Alice implements the encoding measurement $\{M^c\}_c$ but does not send the outcome to the receiver. Instead, she uses an auxiliary maximally entangled state $\ket{\phi^+_{d_C}}$ to implement a dense coding protocol. That is, Alice uses $c$ to select a dense-coding unitary to perform on her half of $\ket{\phi^+_{d_C}}$ and sends the quantum system to the receiver who performs a Bell state measurement to extract $c$. The receiver can then implement the same decoding channels $\Lambda_{c,y}$ as in the original protocol.
\end{proof}
In short, the quantum case can be mapped to the classical case via teleportation and the classical case can be mapped to the quantum case via dense coding. The additional entanglement cost is one auxiliary maximally entangled state $\ket{\phi^+_{d_c}}$. This result may be viewed as a quantum inputs generalisation of the analogous result shown for classical inputs in \cite{Vieira2023}. Note, however, that one would not expect the two resouces to be equivalent if the amount of entanglement allowed in the protocol is restricted.

\subsection{Performance metrics}
While studying the space of correlations $\{\tau_{\psi, y}\}_{\psi,y}$ is the general approach to characterising the PM scenario, it is operationally more natural to consider that Alice and Bob want to implement a specific communication task. This means that when Bob draws $y$, his aim is to learn a specific quantum property of $\psi$. In general, the desired property associated with the input $y$ is described by a channel $\Theta_y$. This means that the goal of Bob is to output the quantum state $\tau_{\psi, y}=\Theta_y(\psi)$. This can be achieved in ideal situations (e.g.~in teleportation) but will typically not be possible. It is therefore relevant to quantify how accurately the states $\tau_{\psi, y}$ approximate the target information $\Theta_y(\psi)$. 

A standard way of quantifying performance is to consider the fidelity between the target $\Theta_y(\psi)$ and the output $\tau_{\psi,y}$. The fidelity between two arbitrary states $\rho$ and $\sigma$ is given by 
\begin{equation}\label{eq:fid_gen}
	F(\sigma,\rho)=\left(\tr\sqrt{\sqrt{\rho}\sigma\sqrt{\rho}}\right)^2.
\end{equation}
When one state ($\sigma$) is pure, which is typically the case since these are often relevant choices of $\Theta_y(\psi)$, the fidelity simplifies to $F(\sigma,\rho)=\bracket{\sigma}{\rho}{\sigma}$. A reasonable quantifier of the performance of a protocol is the fidelity averaged over all the $Y$ decodings and all the quantum inputs $\mathcal{I}$, 
\begin{equation}\label{eq:ave_fid}
	F_\text{avg}= \frac{1}{Y |\mathcal{I}|} \sum_{y=1}^Y\sum_{\psi\in \mathcal{I}} F\left(\tau_{\psi,y},\Theta_y(\psi)\right)
\end{equation} 
where $|\mathcal{I}|$ is the size of the set $\mathcal{I}$. When $\mathcal{I}$ is an uncountable set (e.g.~all pure states of a given dimension) the summation is replaced with an integral. Note that when $Y=1$ and $\Theta_y(\psi)=\psi$, we recover the  average fidelity that is commonly used to benchmark the performance of standard teleportation protocols \cite{Horodecki1999}.

However, the on-average quantifier has the drawback that for some input states the protocol could achieve a fidelity far lower than its average. It is therefore interesting to consider an alternative quantifier based on the worst-case performance. This is the lowest fidelity obtained when optimising it over all input states in $\mathcal{I}$ and over Bob's possible choice of $y$. That is,
\begin{equation}
	F_\text{worst}=\min_{y,\mathcal{I}}  F\left(\tau_{\psi,y},\Theta_y(\psi)\right).
\end{equation}
This is relevant when the dimension of the shared entanglement is restricted, since otherwise the set of correlations becomes convex, thereby allowing the worst-case performance to equal the average performance.

Lastly, we distinguish a particularly powerful type of protocol in which the performance is independent of Alice's and Bob's inputs. In other words, the protocol always achieves the same fidelity regardless of the choice of $\psi$ and $y$, i.e.~
\begin{equation}\label{universal}
	 F_\text{indep}=F\left(\tau_{\psi,y},\Theta_y(\psi)\right), \qquad \forall \psi, y.
\end{equation}
We refer to these protocols as universal when $\mathcal{I}$ corresponds to entire pure quantum state space. As a simple example, the ideal teleportation protocol is universal because it achieves $F_\text{indep}=1$. A similarly spirited example is the universal quantum cloning machine, which achieves a fixed but non-unit fidelity independently of the pure state selected to be cloned \cite{Werner1998}.

\subsection{Numerical method for optimising protocols}\label{search}
In this section we describe a numerical method for  optimising the performance of quantum protocols for a given communication task. For this, we use semidefinite programming (SDP) methods (see the review \cite{Tavakoli2024}) that search the set of quantum correlations from the interior via so-called alternating convex search algorithms.

\subsubsection{Classical messages}
Consider a PM scenario with classical communication. For a given entangled state $\rho_{AB}$, the optimisation of $F_\text{avg}$ is evaluated over Alice's measurement $\{M^c\}_c$ and the set of channels $\{\Lambda_{c,y}\}_{c,y}$ used by Bob to decode his share of $\rho$. Solving this problem exactly is challenging in general, but we now show how to obtain useful lower bounds. To this end, we use state-channel duality to associate each CPTP map $\Lambda_{c,y}^{B\rightarrow B'}$ with a so-called Choi state, $\eta_{c,y} \in \mathcal{D}(\mathcal{H}_{B'}\otimes\mathcal{H}_{B})$. It is well-known that the action of the channel can be expressed as $\Lambda_{c,y}(X) = d_B\tr_B[(\mathds{1}_{B'}\otimes X^T) \eta_{c,y}]$, where $d_B = \dim(\mathcal{H}_B)$ \cite{Skrzypczyk2023}. The Choi state $\eta_{c,y}$ is positive semidefinite and satisfies $\operatorname{tr}_{B'}(\eta_{c,y}) = \mathds{1}/d_B$. Hence, we can represent the correlations as $\tau_{\psi,y} =d_B\sum_{c=1}^{d_C} \tr_B[(\mathds{1}_{B'} \otimes \sigma_{c \mid \psi}^T)\eta_{c,y}]$. When $\Theta_y(\psi)$ are pure, the average fidelity becomes
\begin{equation}\label{eq:ave_fid}
F_\text{avg}= \frac{d_B}{Y |\mathcal{I}|} \sum_{y=1}^Y\sum_{\psi\in \mathcal{I}} \sum_{c=1}^{d_C} \tr[(\Theta_y(\psi) \otimes \sigma_{c \mid \psi}^T)\eta_{c,y}].
\end{equation}
For a given entangled state $\rho$, this expression can be optimised via an alternating convex search algorithm that iterates between an SDP evaluated over $\{M^c\}_c$ and an SDP evaluated over $\{\eta_{c,y}\}_{c,y}$. This iteration can be continued until convergence.  The first sub-routine becomes 
\begin{align}\nonumber
	\max_M & \quad   F_\text{avg}\\
	\text{s.t.}& \quad \sum_{c=1}^{d_C} M^c=\openone_{A'A} \quad \text{and} \quad M^c\succeq 0 \hspace{2mm}\forall c.
\end{align}
The second sub-routine becomes
\begin{align}\nonumber
	\max_\eta & \quad   F_\text{avg}\\
	\text{s.t.}& \quad \tr_{B'}(\eta_{c,y})=\frac{\openone_B}{d_B} \quad \text{and} \quad \eta_{c,y}\succeq 0 \hspace{2mm}\forall c,y.
\end{align}
In scenarios when one also permits the entangled state to be arbitrary, one can add a third sub-routine (also SDP) in which one optimises $F_\text{avg}$ over $\rho_{AB}$ in a selected  a Hilbert space. Moreover, if both $\Theta_y(\psi)$ and $\tau_{\psi,y}$ are mixed, the fidelity can be computed by first purifying $\Theta_y(\psi)$ using Uhlmann's theorem \cite{SkrzypczykBook}.

Furthermore, in order to estimate the worst-case fidelity, $F_{\text{worst}}$, we follow the same approach but change the objective function. We introduce a scalar variable $t$ and impose the semidefinite constraint that $ t \leq F(\tau_{y,\psi}, \Theta_y(\psi))$ for all $y,\psi$. By maximising $t$ in the SDPs, we obtain lower bounds on  $F_{\text{worst}}$.

Lastly, we discuss how to optimise over a relevant universal protocol. Assume that Alice receives $N$ unknown and independent $d$-dimensional pure quantum states, $\ket{\psi} = \bigotimes_{i=1}^N \ket{\psi_i} \in \mathcal{H}_{A'}$. Given that Bob randomly selects $y \in [N]$, he aims to learn the associated state $\psi_y$, i.e., $\Theta_y(\psi) = \psi_y$. In this case, the fidelity can be expressed as
\begin{equation}
F_{y, \psi} =  \tr[(\psi_y \otimes \psi_y) Z_y],
\end{equation}
where the operator $Z_y = d_B \tr_{\bar{A} B}[(\phi^y_{A'} \otimes \rho_{AB}^{T_B} \otimes \openone_{B'})(\sum_c M^c_{A'A}\otimes \eta^{c,y}_{B'B})]$ with $\phi^y = \big( \bigotimes_{i\neq y} \psi_i\big) \otimes \openone_y$. Here, the partial trace goes over all subsystems except $\mathcal{H}_{B'}$ and the $y$'th subsystem of $\mathcal{H}_{A'}$.
For the protocol to achieve the same fidelity regardless of $\psi$ and the input choice $y$, the operator must act identically on any pair of pure states $\ket{\varphi} \in \mathbb{C}^d$. This implies both that $Z_y \equiv Z$ for all $y$ and that $ \tr[(\varphi \otimes \varphi) (Z- \tilde{Z})] = 0$ for all $\varphi$, where $\tilde{Z}$ is the result of twirling $Z$.
The twirling is a symmetrisation operation and yields $\tilde{Z} = \mathcal{O}_{\text{sym}}+\mathcal{O}_{\text{asym}}$, where $\mathcal{O}_{\text{(a)sym}}$ is proportional to the projection onto the (anti-)symmetric subspace of $\mathbb{C}^d \otimes \mathbb{C}^d$. Using that $\tr(\tilde{Z} \, \varphi \otimes \varphi) = \tr(\mathcal{O}_{\text{sym}} \, \varphi \otimes \varphi)$, we find that $Z$ optimally lives in the symmetric subspace. This implies that $\bra{\varphi_{\text{sym}} }Z\ket{\varphi_{\text{sym}}'} = 0$ for all distinct symmetric states $\varphi_{\text{sym}},\varphi_{\text{sym}}'$. Including these constraints in the see-saw algorithm allows us to optimise over universal protocols. In particular, for qubit states $\varphi$ we can w.l.g. impose that $Z = (1-a)\,\openone + (2a-1)\,\text{SWAP}$ for $a \in [0,1]$.

\subsubsection{Quantum messages}
The above algorithm can be adapted to address also the case of quantum messages. To this end, one needs to apply the state-channel duality also to the quantum encoding channels $\Gamma^{A'A\rightarrow C}$ to represent them in terms of Choi states, $\mu  \in \mathcal{D}(\mathcal{H}_{A'}\otimes \mathcal{H}_{A}\otimes \mathcal{H}_{C})$, where $\mathcal{H}_{C}$ represents the $d_C$-dimensional Hilbert space of the quantum message. The total state of Bob prepared by Alice then reads 
\begin{equation}
\sigma_{\psi} = d_{A'}d_{A}\tr_{A'A}[(\mathds{1}_C \otimes (\psi_{A'} \otimes \rho_{AB}) ^{T_{A'A}})(\mu \otimes \mathds{1}_B)] \,
\end{equation}
where $T_{A'A}$ denotes partial transposition over the $A'A$ system and
 $d_{A'} = \dim(\mathcal{H}_{A'})$ and $d_{A} = \dim(\mathcal{H}_{A})$. The average fidelity is then given by
\begin{equation}
F_\text{avg}= \frac{d_B d_C }{Y |\mathcal{I}|} \sum_{y=1}^Y\sum_{\psi\in \mathcal{I}} \tr[(\Theta_y(\psi) \otimes \sigma_{\psi}^T)\eta_{y}].
\end{equation}
where Bob's Choi state now acts on $\eta_y \in \mathcal{D}(\mathcal{H}_{B}\otimes \mathcal{H}_{C}\otimes \mathcal{H}_{B'})$. This again leads to an alternating search that iterates between two SDP sub-routines, but now with the former routine evaluating over the Choi states of Alice. These states are characterised by $\mu \succeq 0$ and $\tr_{C}(\mu) = \frac{\mathds{1}_{A' A}}{d_{A'}d_A}$. In the second sub-routine, Bob's Choi states are characterised by $\eta_{y}\succeq 0$ and $\tr_{B'}(\eta_y)=\frac{\mathds{1}_{BC}}{d_B d_C}$.

\subsubsection{Unlimited pure input states}
The above search methods can be applied when $\mathcal{I}$ is a finite set, but it is often relevant to consider $\mathcal{I}$ as uncountably infinite. Of particular importance is the case where $\mathcal{I}$ corresponds to all pure quantum state of given dimension. To deal with this case, we use spherical designs. 

Specifically, assume that Alice receives an arbitrary pure quantum state $\psi \in 
\mathcal{P}(\mathbb{C}^d) = \{ |\psi\rangle \in \mathbb{C}^d \mid \langle \psi | \psi \rangle = 1 \}$. The average fidelity in the PM scenario then reads \eqref{eq:ave_fid}
\begin{equation}\label{eq:fidelity_int}
    F_{\text{avg}} = \frac{1}{Y} \sum_{y=1}^Y \int_{\mathcal{P}(\mathbb{C}^d)} \ \  \text{d}  \psi \ F(\tau_{\psi, y}, \Theta_y(\psi)) \,, 
\end{equation}
where the integral is taken over all pure $d$-dimensional states with respect to the Haar measure. To deal with this expression, we note that the fidelity is a second-order polynomial in $\psi$. This follows from the fact that every quantum channel is linear map, which guarantees that both the output state $\tau_{\psi, y}$ and the target state $\Theta_y(\psi)$ are polynomials of degree one in $\psi$. Consequently, the Haar average of the fidelity can be substituted with a spherical $2$-designs. In general, a set $\{|\phi_k\rangle\}_{k=1}^K$ of $K$ normalized vectors $\phi_k \in \mathcal{P}(\mathbb{C}^d)$ is a spherical $t$-design if and only if the average value of any $t$-th order polynomial $p_t(\psi)$  over the set $\{|\phi_k\rangle\}$ is equal to the average of $p_t(\psi)$ over all $\psi \in \mathcal{P}(\mathbb{C}^d)$ \cite{Renes2004}. We can thus express the average fidelity as the finite sum
\begin{equation}\label{eq:fid_t_des}
    F_{\text{avg}} = \frac{1}{YK}\sum_{y=1}^Y\sum_{k = 1}^K  F(\tau_{\phi_k, y}, \Theta_y(\phi_k)) \,,
\end{equation}
where the set $\{|\phi_k\rangle\}$ forms a 2-design in dimension $d$. 

The design can be chosen freely. A simple and systematic choice is a symmetric informationally complete (SIC) POVM \cite{Renes2004}. This is a set of $d^2$ equiangular vectors that resolve the identity. The construction of SIC POVMs strongly relies on the Weyl-Heisenberg (WH) group. The WH group has two generators, which can be chosen as the so-called shift and clock operators $X = \sum_{j=0}^{d-1} |j\rangle \langle j-1|$ and $Z = \sum_{j=0}^{d-1} \omega^j | j\rangle \langle j|$, respectively, where $\omega = e^{2\pi i/d}$. Every known SIC POVM (with a single exception for dimension eight) is generated as the orbit of the WH group $|\phi_k\rangle = X^{k_0}Z^{k_1} |\varphi\rangle$ for $k = (k_0,k_1) \in [d^2]$ for a suitable fiducial state $|\varphi\rangle$ \cite{Scott2010}.


\section{Stochastic teleportation}
 
\begin{figure}[t!]
	\centering
	\includegraphics[width=\columnwidth]{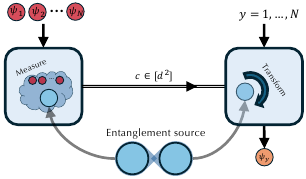}
	\caption{\textit{Stochastic teleportation.} Alice receives $N$ $d$-dimensional states and Bob aims to recover the $y$'th state. As in standard teleportation, the parties can use a shared source of entangled particles and classical messages of size $d^2$.}\label{Fig_stoch_teleport}
\end{figure}

We now apply the framework for quantum inputs in the PM scenario to investigate  quantum communication tasks that naturally extend  quantum teleportation. We call this class of tasks, of which standard teleportation is the simplest example, \textit{stochastic teleportation}. In the stochastic teleportation protocol, Alice receives $N$ unknown, independent and randomly selected, $d$-dimensional pure quantum states,
\begin{equation}
	\ket{\psi}\equiv \ket{\psi_1}\otimes\ket{\psi_2}\otimes\ldots\otimes \ket{\psi_N} \in \mathcal{H}_{A'} \,.
\end{equation}
Thus, Alice's state space corresponds to $\mathcal{I}=\mathcal{P}(\mathbb{C}^d)^{\otimes{N}}$ and the dimension of her input state $\psi$ is $\text{dim}(\mathcal{H}_{A'})=d^N$. Bob privately selects a symbol $y\in[N]$, which indexes the state $\ket{\psi_y}$ he wishes to learn. Hence, the CPTP maps $\Theta_y$, which describe the targeted quantum information, are partial-trace maps,
\begin{equation}
	\Theta_y(\psi)=\tr_{\neg y}(\psi)=\psi_y \, ,
\end{equation} 
where  $\neg y$ indicates that the partial-trace goes over the sub-systems except $y$, i.e.~$\{1,\ldots,N\}\setminus\{y\}$, of $\mathcal{H}_{A'}$. Next, we select the communication resources allowed for Alice and Bob. In addition to having a shared state, we allow them to use two dits of classical communication, corresponding to $c\in[d^2]$. This choice is motivated by standard teleportation, which we now recover by selecting $N=1$. For $N >1$ Alice's task is, without using more communication, to teleport a state to Bob from which he can extract any one of Alice's $N$ separate states; see Fig~\ref{Fig_stoch_teleport}. The average fidelity of stochastic teleportation therefore becomes
\begin{equation}
	F_\text{avg}=\frac{1}{N}\sum_{y=1}^N \int_{\mathcal{P}(\mathbb{C}^d)} \ \text{d} \psi \ \bracket{\psi_y}{\tau_{\psi,y}}{\psi_y} \,.
\end{equation}

An important fact is that stochastic teleportation of $N>1$ qudits cannot be performed perfectly in quantum theory unless Alice is allowed to communicate as much classical information as would be required for the standard teleportation of the $N$ qudits. The next result proves this no-go statement.

\begin{result}[Impossibility of perfect stochastic teleportation]\label{no-go}
Stochastic teleportation of $N$-inputs of dimension $d$ cannot be achieved with unit fidelity unless $2 N \log d$ bits of classical communication are permitted. 
\end{result}
\begin{proof}
We present a proof by contradiction. Specifically, we show that \emph{perfect stochastic teleportation} with $M < 2N \log d $ bits of classical communication violates information causality \cite{Pawlowski2009}. Consider that Alice holds the data string $x = x_1 \dots x_N$, where each element $x_k=(u_k,v_k) \in [d^2]$. Bob privately and uniformly selects $y \in [N]$, with the goal of outputting $x_y$. 
To achieve this task, Alice and Bob share $N$ copies of the maximally entangled state,  $(\phi^+_d)_{A_1B_1} \otimes \ldots\otimes (\phi^+_d)_{A_NB_N}$, where $\ket{\phi^+_d}_{A_k B_k} = \frac{1}{\sqrt{d}}\sum_{i=0}^{d-1}|ii\rangle$ for all $k$. For each $k \in [N]$, Alice  encodes the element $x_k = (u_k, v_k)$, by performing the unitary $U_{x_k} = X^{u_k} Z^{v_k}$ on her share of $(\phi^+_d)_{A_kB_k}$. 

By assumption, the parties can now implement a perfect stochstic teleportation protocol, which allows Bob to recover any $y \in [N]$ of Alice's qudits perfectly while consuming only $M < 2N \log d$ bits of classical communication. He then performs a Bell-state measurement jointly on the $y$'th qudit obtained from the stochastic teleportation protocol and his share of the corresponding entangled state $(\phi^+_d)_{A_yB_y}$. This corresponds precisely to a dense coding protocol, yielding the outcome $x_y$.
Hence, Bob can recover any element $x_y$ in Alice data $x$ perfectly.

This procedure implements a perfect Random Access Code (RAC), where a string of length $N$ where each element takes $d^2$ possible values is stochastically communicated with less than $M < 2N \log d$ bits of classical communication. This violates information causality \cite{Pawlowski2009}. Consequently, we arrive at a contradiction, which implies that perfect stochastic teleportation is impossible with less resources required for standard teleportation.
\end{proof}

Even though perfect stochastic teleportation is impossible, with the same amount of classical communication as in standard teleportation, it may still be possible to perform the task at with high fidelity. In what follows, we discuss how such protocols can be developed.

\subsection{Protocols based on teleportation and random access coding}
\begin{figure}[t!]
    \centering
    \begin{subfigure}[b]{0.45\textwidth}
        \includegraphics[width=\textwidth]{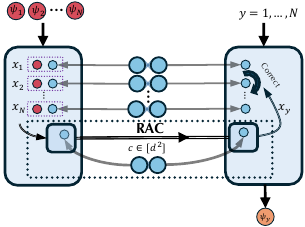}
        \caption{}
    \end{subfigure}
    \hfill
    \begin{subfigure}[b]{0.45\textwidth}
        \includegraphics[width=\textwidth]{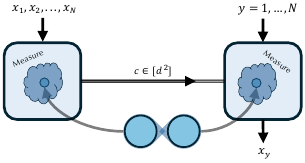}
        \caption{}   
    \end{subfigure}
    \caption{\textbf{(a)} \emph{Stochastic teleportation using Random Access Codes.} Alice independently performs a Bell state measurement on each of her input qudits $\psi_x$ with half of a shared $\phi_d^+$ state, obtaining the outcomes  $x = (x_1, \dots, x_N)$. The parties use a Random Access Coding protocol, allowing Bob to retrieve the specific outcome  $x_y$, which determines the correction  $U_{x_y}$  he applies to the  $y$'th qudit.
    \textbf{(b)} \emph{Detail of the Random Access Code}. The parties implement an entanglement-assisted Random Access Code, using $ d^2$ classical messages and shared entanglement to stochastically transmit one of the  $d^2$-valued symbols $x_y$ of Bob's choice from Alice's bitstring $x$.}\label{fig:stoch_tele_rac}
\end{figure}

We begin with discussing how the combination of two quantum information primitives, namely standard teleportation and random access codes (RACs), can be used to perform stochastic teleportation. For this purpose, consider a protocol of the following form (see Fig~\ref{fig:stoch_tele_rac}):
\begin{enumerate}
	\item Alice and Bob share $N$ pairs of the maximally entangled state, $(\phi^+_d)_{A_1B_1} \otimes \ldots\otimes (\phi^+_d)_{A_NB_N}$, where $\ket{\phi^+_d}_{A_k B_k} = \frac{1}{\sqrt{d}}\sum_{i=0}^{d-1}|ii\rangle$ for all $k$. 
	\item For each $k\in[N]$, Alice performs a complete Bell state measurement on $\psi_k$ and her share of $(\phi^+_d)_{A_kB_k}$. This measurement is defined as the basis $\{U_{x_k} \otimes \openone_{A} \ket{\phi^+_d}_{A'A}\}_{x_k=1}^{d^2}$, where the unitary $U_{x_k}=X^{u_k}Z^{v_k}$ is associated with the measurement outcome $x_k=u_kv_k\in\{0,\ldots,d-1\}^2$. We denote Alice's complete set of outcomes by $x=x_1\ldots x_N$ and note that it is uniformly random. After Alice's measurements the state of Bob's $k$'th particle (system $B_k$) is given by $U_{x_k}^\dagger \psi_k U_{x_k}$. 
	\item In order to recover an accurate copy of $\psi_y$ corresponding to Bob's choice $y$, Bob must learn the classical data $x_y$, which informs the correction unitary $U_{x_y}$ on his share of $(\phi^+_d)_{A_yB_y}$. To execute this communication step, Alice must encode $x$ into the classical message $c\in[d^2]$ so that Bob can decode any data element $x_y$, given that he privately selects $y \in [N]$. In the literature, this well-established task is known as a RAC.
\end{enumerate}

The described protocol boils stochastic teleportation down to performing a RAC. The RAC task takes place in an PM scenario with \textit{classical} inputs. Alice holds the data string $x=x_1\ldots x_N$ where $x_i\in[d^2]$ for $i=1,\ldots,N$, which she encodes into a message that is sent to Bob. Bob draws the variable $y\in[N]$ and from reading Alice's message, he must then output the classical data $b=x_y$, where $b\in [d^2]$. When the input data $(x,y)$ is uniformly distributed, the average success probability of the RAC is
\begin{equation}
	P_\text{RAC} = \frac{1}{N d^{2N}}\sum_{x,y}p(b = x_y|x,y).
\end{equation}
This type of task has been used in many quantum information contexts, e.g.~fundamental principles for quantum nonlocality \cite{Pawlowski2009}, quantum cryptography \cite{Chailloux2016} and winning competitive card games \cite{Sadiq2014}.

It stands to reason that if one can perform a perfect RAC, namely $P_\text{RAC}=1$, then Bob can always recover $x_y$. Then, via the third step of the above protocol, he can deterministically output $\tau_{\psi,y}=\psi_y$ and thereby succeed the stochastic teleportation task with unit probability ($F_\text{avg}=1$). Below, we show that this connection is quantitative also when the RAC cannot be performed at perfect success rate. 

\begin{result}[Connection to random access codes]\label{res:Ave_EARAC}
For every protocol that realises a RAC at success rate $P_\text{RAC}$ for $N$ separate $d^2$-valued inputs, there exists a corresponding $N$-input $d$-dimensional stochastic teleportation protocol with average fidelity
\begin{equation}
	F_\text{avg}= \frac{d P_\text{RAC} +1}{d+1} \,.
\end{equation}
\end{result}

\begin{proof}
Consider the stochastic teleportation protocol described above in which Bob's final guess for $|\psi_y\rangle$ is given by $|\tau_{\psi,y}\rangle = U_b U_{x_y}^\dagger |\psi_y\rangle$. Now, with probability $P_\text{RAC}$ the RAC succeeds and Bob performs the correct unitary, yielding  $\tau_{\psi,y} = \psi_y$. With probability $1- P_\text{RAC}$ the RAC fails and Bob applies one of the $d^2-1$ incorrect rotation to his $y$'th state. By invoking shared randomness between Alice and Bob, the probability of the $d^2-1$ failing outcomes in the RAC can be taken as uniform. Moreover, the total rotation $U_b U_{x_y}^\dagger $ corresponds to an operator of the form  $V_{ij} = X^{i} Z^j$, for $i,j\in\{0,\ldots,d-1\}$. Bob's average output state $\tau_{\psi,y}$ is thus given by 
\begin{equation}
\begin{aligned}
    \tau_{y,\psi} &= P_\text{RAC}\psi_y + \frac{1-P_\text{RAC}}{d^2-1}\sum_{\substack{i,j=0 \\ (i,j)\neq (0,0)}}^{d-1} V_{ij} \psi_y  V_{ij}^\dagger\\
			&=  \bigg(P_\text{RAC} - \frac{1-P_\text{RAC}}{d^2-1}\bigg) \psi_y + \frac{1-P_\text{RAC}}{d^2-1}\sum_{i,j=0 }^{d-1} V_{ij} \psi_y  V_{ij}^\dagger  \,.
\end{aligned}
\end{equation}
In the second row we have added and subtracted the term $ \frac{1-P_\text{RAC}}{d^2-1}\psi_y$, such that the sum in the second term corresponds to an unormalised Weyl-twirling of the state $\psi_y$, meaning that $\sum_{i,j=0 }^{d-1} V_{ij} \psi_y  V_{ij}^\dagger = d \openone$ \cite{Wilde2017}. Hence, the fidelity of the output state simply becomes $\langle \psi_y |\tau_{\psi,y} |\psi_y\rangle = \frac{dP_\text{RAC}+1}{d+1}$ for all $\psi,y$.
\end{proof}

Result~\ref{res:Ave_EARAC} establishes a one-to-one correspondence between the average success rate of stochastic teleportation and RACs. This enables us to build on the previous literature on RACs to optimise the performance of stochastic teleportation.

\subsection{Perfect stochastic teleportation with no-signalling boxes}
Before considering quantum protocols, we begin by considering how well stochastic teleportation can be implemented with post-quantum resources.  Following Result~\ref{res:Ave_EARAC}, we focus on evaluating the performance of a RAC when the shared resource between Alice and Bob supports post-quantum nonlocality, often referred to as nonlocal boxes. In \cite{Grudka2015}, it was shown that if Alice and Bob share a Popescu-Rohrlich box \cite{Popescu1994}, then the simplest stochastic teleportation protocol, in which the inputs are a pair of qubits [$(N,d)=(2,2)$], can be performed perfectly. Here, we show that such a perfect performance is possible for any $(N,d)$, if one uses a more general nonlocal box.

In general a nonlocal box is any bipartite probability distribution $p(a,b|x,y)$ that satisfies the no-signaling principle, $p(a|x,y)=\sum_b p(a,b|x,y)=p_A(a|x)$ and  $p(b|x,y)=\sum_a p(a,b|x,y)=p_B(b|y)$, and it is well-known that these often are not realisable in quantum theory \cite{Barrett2005}. To use these to boost the performance of the RAC, we first note that specific Bell inequalities can be translated into the success probability of a corresponding RAC \cite{Tavakoli2016}. Then, we construct a no-signaling box for the Bell inequality.

Let Alice use $x$ as an input to a Bell inequality test and denote her outcome by  $a\in[d^2]$. Bob uses $y$ as an input for the Bell inequality test and outputs $b\in[d^2]$. Alice then  sends $a$ to Bob who outputs $a\oplus b$ as his guess for $x_y$. Here, $\oplus$ denotes addition modulo $d^2$. In this approach, the average success probability of the RAC equates with the following Bell parameter \cite{Tavakoli2016}
\begin{equation}\label{PBell}
P_\text{Bell} = \frac{1}{N d^{2N}}\sum_{x,y} p(a\oplus b = x_y|x,y) \,.
\end{equation}
Hence, this strategy gives $P_\text{RAC}=P_\text{Bell}$.  We now show that there exists a no-signaling box that achieves the maximal value, $P_\text{Bell}=1$.

\begin{result}[Nonlocal boxes trivialise stochastic teleportation] \label{Result3}
The $N$-input and $d$-dimensional stochastic teleportation task can be performed perfectly if the parties share a nonlocal box and send two dits of classical communication. This holds independently of the number of quantum inputs $N$.
\end{result}

\begin{proof}
	Using Result~\ref{res:Ave_EARAC}, it suffices to show that one can achieve $P_\text{RAC}=1$. Via the above strategy that equates $P_\text{RAC}$ with a corresponding Bell parameter $P_\text{Bell}$, it suffices to show that there exists a nonlocal box that achieves $P_\text{Bell}=1$. To that end, consider the following bipartite probability distribution 
	\begin{equation}\label{eq:NS_C}
	p(a,b|x,y)= \begin{cases} \frac{1}{d^2}  & a\oplus b  = x_y \\ 0 & \text{otherwise}.\end{cases} 
	\end{equation}
	This distribution is non-negative, normalised and no-signaling; the marginals are $p_A(a|x)=p_B(b|y)=\frac{1}{d^2}$. When inserted in \eqref{PBell}, it achieves $P_\text{Bell}=1$.
\end{proof}
The fact that Result~\ref{Result3} is independent of $N$ means that using only a constant amount of classical communication, an arbitrary number ($N$) of quantum inputs can be stochastically teleported to Bob with unit fidelity. This may be viewed as a manifestation of the exceptional capabilities of post-quantum nonlocality. It is known that nonlocal boxes can trivialise communication complexity for classical inputs \cite{Brassard2006}. Result~\ref{Result3} can be interpreted as the analogous phenomenon for quantum inputs.

\subsection{Entanglement-assisted protocols}
A quantum protocol can use entanglement between Alice and Bob in order to efficiently perform the RAC, which via Result~\ref{res:Ave_EARAC} leads to a corresponding stochastic teleportation fidelity. We now analyse such protocols. The parties can share an entangled state $\rho_{AB}$. For each $x$, Alice performs an associated measurement $\{A_{c\mid x}\}_c$ on her share of the entangled state, where the measurement outcome $c \in [d^2]$ corresponds to the classical message sent to Bob. Bob waits until he receives $c$, and thereafter uses this information together with his random input $y \in [N]$ to select the quantum measurement $\{B_{b \mid y,c}\}_b$, where the output $b \in [d^2]$ is his final guess for $x_y$. The resulting probabilities are given by Born's rule 
\begin{equation}
p(b | x,y) = \sum_{c=1}^{d^2} \tr(A_{c \mid x} \otimes B_{b \mid y, c} \ \rho_{AB}) \,.
\end{equation}
Thus, the success probability in the random access code becomes
\begin{equation}\label{EARAC}
P_\text{RAC}=\frac{1}{N d^{2N}}\sum_{x,y}\sum_{c=1}^{d^2} \tr(A_{c \mid x} \otimes B_{x_y| y, c} \ \rho_{AB}) \,.
\end{equation}

A natural way to approach the task of constructing relevant protocols is to let  Alice and Bob share the maximally entangled state $\ket{\phi^+_{d^2}} $, with Alice encoding her message via rank-1 projective measurements. For any such protocol, we derive an upper bound on the optimal average success probability.

\begin{result}[RAC with maximally entangled state] \label{res:EARAC_bound}
    Consider a random access code assisted by a maximally entangled state $\phi^+_{d^2}$ and rank-1 projective encoding measurements. In the setting of $N$-element data with alphabet size $d^2$ and classical message dimension $d^2$, the average success probability is bounded by
    \begin{equation} \label{eq:Ana_EARAC}
    P_\text{RAC}\leq \frac{1}{N}\left(1+\frac{N-1}{d}\right) \,.
    \end{equation}
\end{result}

\begin{proof}
We transform the entanglement-assisted RAC to a PM scenario without entanglement but based on sending quantum messages (QRAC). This is possible \cite{Zukowski2017} because thanks to the selected state and the rank-1 projective measurements  Alice's outcomes are uniformly random. In the corresponding PM scenario, Alice has two inputs $(c,x)$ and sends the (normalised) states that she would have prepared remotely for Bob in the entanglement-assisted scenario, namely $\sigma_{c\mid x} = d^2\operatorname{Tr}_A[(A_{c\mid x} \otimes \mathds{1}) \phi^+] \otimes |c\rangle \langle c|$. Once Bob receives the states he reads the classical register and performs an associated measurement $\{B_{b \mid y,c}\}$ with outcome $b$ on the quantum register. We then have that 
\begin{align}\nonumber
	P_{\text{RAC}}&=\frac{1}{N d^{2N+2}}\sum_{x,y,c}\tr\left(\sigma_{c|x} B_{x_y|y,c}\right)\\\nonumber
	& \leq \frac{1}{N d^{2N}} \sum_{x,y} \max_c \tr\left(\sigma_{c|x} B_{x_y|y,c}\right)\\\label{qrac}
	& \leq \frac{1}{Nd^{2N}} \sum_x \| \sum_y B_{x_y\mid y} \|_\infty,
\end{align}
where we have defined  $B_{x_y|y}$ as the $B_{x_y|y,c}$ associated with the optimal value of $c$. In the last step, we have relaxed the no-signaling condition on the states and bounded $P_{\text{RAC}}$ over the set of all states. The expression \eqref{qrac} is precisely the expression obtained when optimising QRACs and a generic bound on this quantity is derived in Result 2 of Ref.~\cite{Farkas2025}. Using this bound gives Eq~\eqref{eq:Ana_EARAC}.
\end{proof}


The bound \eqref{eq:Ana_EARAC} provides a simple limitation on natural classes of quantum protocols, but it is not expected to be tight in general. From Result~\ref{res:Ave_EARAC}, we obtain the corresponding bound on the average stochastic teleportation fidelity,
\begin{equation}
F_\text{avg}\leq \frac{2N+d-1}{N(d+1)} \,.
\end{equation}

The simplest interesting for stochastic teleportation scenario concerns a pair of qubits, i.e.~$(N,d)=(2,2)$. We have gone beyond the above type of quantum strategy and considered the random access code value under the most general quantum protocol, as described in \eqref{EARAC}, which could use arbitrary measurements and potentially unbounded entanglement. Our next result proves the optimal value for this scenario.
\begin{result}[Optimal RAC for simplest case]
When $N=d=2$, the optimal quantum protocol for the random access code defined in Eq.~\eqref{EARAC} achieves
\begin{equation}
P_\text{RAC}=\frac{3}{4}.
\end{equation}
\end{result}
\begin{proof}
First we used the numerical search method outlined in section~\ref{search} to find an explicit quantum protocol that achieves $P_\text{RAC}=3/4$ up to numerical precision. This protocol is based on using a four-dimensional maximally entangled state and it is adaptive in the sense of Ref.~\cite{Pauwels2022}, i.e.~Bob must wait to receive the message before selecting his measurement. While we have not found an analytical form, we provide all the measurements of Alice and Bob in an open repository file \cite{code}.

Next, we prove that no quantum protocol can exceed  $P_\text{RAC} = 3/4$. To establish this, we use the framework of informationally restricted correlations \cite{info1, Chaturvedi2020}, which characterizes the correlations attainable when Alice and Bob have classical inputs and communication is constrained by the entropic content of Alice's messages. It was shown in \cite{Tavakoli2021} that quantum correlations in entanglement-assisted PM scenarios with classical inputs and classical communication can be upper bounded by the correlations achievable without entanglement but with quantum messages subject to specific informational restrictions. Leveraging this result, we apply the hierarchy of semidefinite relaxations developed in Ref.~\cite{info2} to bound  $P_\text{RAC}$. The resulting SDP is based on the positivity of a large matrix; in order to handle it we employ techniques from SDP symmetrisation \cite{Rosset2019, Ioannou2022}. Our computations confirm  $P_\text{RAC} = 3/4$  up to solver precision.
\end{proof}
Via Result~\ref{res:Ave_EARAC}, it follows that for a pair of qubits to be stochastically teleported, the optimal quantum protocol based on the RAC achieves $F_\text{avg}=\frac{5}{6}$.


\section{Universal stochastic teleportation}
So far, we have focused on protocols designed to perform stochastic teleportation with high average fidelity. However, protocols based on random access codes (RACs) tend to consume a significant amount of entanglement. For instance, in the simplest case of two qubits ($N = d = 2$), our RAC-based protocol requires four ebits: two copies of $\phi^+_2$ as teleportation building blocks and one copy of $\phi^+_4$ for the RAC component.

In this section, we introduce alternative protocols that are based neither on standard teleportation nor on RACs. These new protocols aim for a stronger benchmark: universal stochastic teleportation. As discussed around Eq.~\eqref{universal}, this means that the teleportation fidelity is independent of both Alice's quantum input and Bob's choice of $y$.

For the case $N = d = 2$, we present an analytical protocol that matches the fidelity of the RAC-based protocol while requiring only one ebit of entanglement and achieving this fidelity uniformly across all inputs. The key to this improvement is that Alice performs genuinely three-particle entangled measurements. The next result demonstrates how this is achieved.

\begin{result}[Universal stochastic teleportation]\label{Res:uni_fid}
For a pair of qubits ($N=d=2$) there exists a stochastic teleportation protocol that consumes one ebit and achieves the fidelity
    \begin{equation}\label{fidelity}
        F_{\text{indep}} = \frac{5}{6} \,.
    \end{equation}
for either qubit, for any pair of pure quantum inputs.  
\end{result}

\begin{proof}
We outline the main steps but defer the complete derivation to Appendix~\ref{App:Worst_case_fid}. Alice starts by performing the joint multi-particle measurement $\{M^c\}_c$, with outcome $c = c_0c_1\in \{0,1\}^2$, on the three-qubit system consisting of her two input qubits $\ket{\psi_1}, \ket{\psi_2}$ and her share of the entangled state $|\phi_2^+\rangle = \frac{1}{\sqrt{2}}(|00\rangle + |11\rangle)$. Each measurement is constructed as $M^{c} = \sum_{k= 0,1} |\psi_{c_0c_1}^k\rangle \langle \psi_{c_0c_1}^k|$, where the set of pairwise orthonormal  states $ |\psi_{c_0c_1}^k\rangle$ is given by
\begin{equation}
    |\psi^k_{c_0c_1}\rangle = X^{c_1+c_0+k}Z^{c_1} \otimes X^{c_1+k}Z^{c_0} \otimes X^k |\psi^{0}_{00}\rangle \, .
\end{equation}
The structure of the fiducial state, $\ket{\psi_0^{00}}$, is inspired by the unitary transformation in universal quantum cloning, see for example Ref.~\cite{Buzek1996}. It takes the form
\begin{equation}
|\psi_{00}^0 \rangle = \sqrt{\frac{2}{3}} |00\rangle_{A'}|1\rangle_{A} - \frac{1}{\sqrt{3}} |\psi^+\rangle_{A'} |0\rangle_{A} \,,
\end{equation}
where $|\psi^+\rangle = \frac{1}{\sqrt{2}}(|01\rangle + |10\rangle)$. After Alice's measurement she sends the 2-bits message $c$ to Bob. Based on the message and his input $y\in [2]$, Bob performs an associated unitary transformation $U^{c,y}$ on his share of the maximally entangled state, given by
\begin{equation}
U^{c,y} =  X^{1+yc_0+c_1}Z^{1+(1+y)c_0+yc_1} \,.
\end{equation}
As shown in Appendix~\ref{App:Worst_case_fid}, the final state of Bob's qubit when averaged over Alice's possible measurement outcomes becomes
\begin{equation}
\tau_{y,\psi} = \frac{5}{6} \psi_y + \frac{1}{6}\psi_y^\perp \,,
\end{equation}
where $|\psi_y^\perp\rangle$ is the state orthogonal to $|\psi_y\rangle$. This is independent of Alice's input state and leads to the fidelity in \eqref{fidelity}.
\end{proof}


It is worth noting that via Result~\ref{Result1}, we can replace the two bits of classical communication with a single qubit and achieve the same fidelity, at the cost of using one additional ebit for a dense coding sub-routine. As a natural benchmark for the protocol, we consider case where the EPR state is subjected to isotropic noise: $\rho_v = v \phi^+ + \frac{1-v}{4}\openone$, where the visibility parameter $v \in [0,1]$. Then we have a quantum-over-classical advantage for $v > 1/2$. This follows form the fact the classical protocol, in which we have with no entangled resources, can achieve up to $F_{\text{avg}} = 2/3$. Furthermore, since universal stochastic teleportation is a linear operation it applies not only to pure input states but also to mixed input states and to inputs that are part of an entangled state. In the following, we consider the application of our protocols also to these cases.

\subsection{Beyond pure inputs}
Let us revisit the task of stochastic teleportation for two qubits, $\rho = \rho_1 \otimes \rho_2 \in \mathcal{H}_{A'}$. As before, the states $\rho_1$ and $\rho_2$ are chosen at random, but we now allow them to be mixed, with purity $T = \tr(\rho_y^2) \leq 1$. Our goal is to determine the fidelity of the protocol as a function of $T$.

We begin by observing that any generic mixed qubit state can be diagonalized in the form $\rho_y = \lambda \psi_y + (1 - \lambda) \psi_y^\perp$, where $\psi_y^\perp$ is orthogonal to $\psi_y$ and $\lambda \in [\frac{1}{2}, 1]$. Due to the linearity of the universal stochastic teleportation protocol, Bob's output state takes the form:
\[
\tau_{\rho, y} = \left( \frac{4}{3}\lambda - \frac{2}{3} \right)\psi_y + \left( \frac{5}{6} - \frac{2}{3}\lambda \right)\openone.
\]
The fidelity between Bob's output $\tau_{\rho, y}$ and the original mixed state $\rho_y$, for $y \in [2]$, is then given by
\begin{equation}
F(T) = \frac{1}{6} \left( 1 + 4T + \sqrt{2} \sqrt{8T^2 - 21T + 13} \right),
\end{equation}
where we have used the relation between purity and $\lambda$, namely $T = 2\lambda^2 - 2\lambda + 1$.
Since the fidelity function is concave, we find that $F(T)$ decreases monotonically with $T$. In particular, for pure states ($T = 1$), we recover $F(1) = \frac{5}{6}$, while for maximally mixed states ($T = \frac{1}{2}$), the fidelity reaches $F(\frac{1}{2}) = 1$.
 
Another relevant question is how well our universal stochastic teleportation protocol performs at entanglement swapping. Consider two input qubits $\psi = (\psi_\theta)_{A_1' C_1} \otimes (\psi_\theta)_{A_2' C_2}$, each part of a partially entangled state $\ket{\psi_\theta}_{A_y' C_y} = \cos \theta \ket{00} + \sin \theta \ket{11}$, with a fixed entanglement parameter $\theta \in [0, \pi/4]$. Alice receives the two qubits $A_1'$ and $A_2'$ as her input.
Suppose Bob draws $y \in [2]$ and aims to reconstruct the joint state $\tau_{\psi,y}$, composed of his share and system $C_y$, such that it approximates $\psi_\theta$ with high fidelity. To achieve this, the parties execute the universal stochastic teleportation protocol.

Crucially, since each of the three particles that Alice holds is entangled with an external partner, her genuine multipartite entangled measurement $M^c \in \mathcal{H}_{A'} \otimes \mathcal{H}_A$ collapses the state of the remaining three particles in $\mathcal{H}_C \otimes \mathcal{H}_B$ into an entangled state. That is, Alice's measurement effectively entangles both qubits $C_1$ and $C_2$ with Bob's part of the maximally entangled state.

After Bob performs the local unitary correction, the total state is given by
$\tau_{y,\psi}  = \frac{2}{3} \psi_\theta+ \frac{1}{6}(\cos^2 \theta|00\rangle\langle 00| + \sin^2 \theta |11\rangle \langle 11 |) + \frac{1}{6}(\cos^2 \theta |01\rangle \langle 01|+  \sin^2 \theta |10\rangle \langle 10|)$.
The entanglement fidelity as a function of $\theta$ is then given by
\begin{equation}
F(\theta) = \frac{5}{6} - \frac{\sin^2(2\theta)}{12}.
\end{equation}
This fidelity reaches its minimum for maximally entangled inputs, yielding $F\left( \frac{\pi}{4} \right) = \frac{3}{4}$.

\subsection{Beyond the simplest scenario}

We now apply the numerical methods described in Section~\ref{search} to investigate universal stochastic teleportation protocols beyond the simplest setting.

\begin{figure}
\centering
\includegraphics[width=0.4\textwidth]{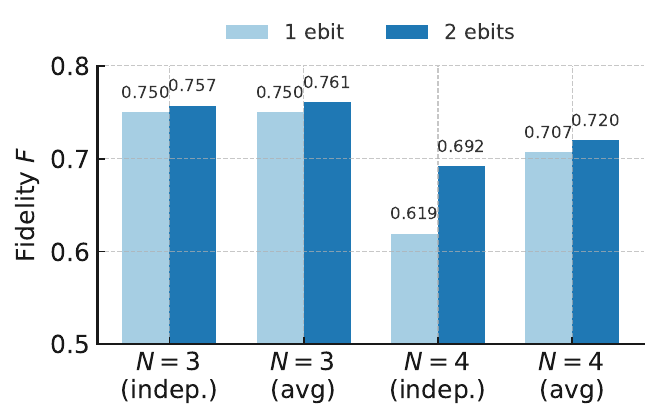}
\caption{Lower bounds on the state-independent and average fidelity for $N = 3,4$ input qubits as a function of entanglement resources.}\label{fig:inc_ebit}
\end{figure}

We begin by analysing qubit protocols with $N = 3$ and $N = 4$, where classical communication is limited to two bits. For each case, we numerically optimise both state-independent fidelity and the average fidelity, considering scenarios in which the parties share one or two ebits. The results, summarised in Fig.~\ref{fig:inc_ebit}, show that for $N = 3$, a universal stochastic teleportation protocol exists that consumes a single ebit and achieves a state-independent fidelity of $F_{\text{indep}} = 3/4$, which matches the corresponding optimal average fidelity. Moreover, we note that performance of the protocols increases with the local dimension of the entangled state, and that this improvement is more pronounced when the number of input qubits is larger. This means that the optimal stochastic teleportation protocols rely on quantum compression operations in the decoding step, which is a phenomenon that has previously been observed for encoding procedures in the context of prepare-and-measure scenarios with classical information \cite{Guo2025}. This motivates an interesting question—reminiscent of an open problem in the standard prepare-and-measure scenario \cite{Pauwels2022b}—is whether there exists an upper bound on the local dimension of the entangled state that is useful for universal stochastic teleportation. We further find that no other shared entangled state can yield a higher fidelity than the maximally entangled state.



Lastly, we have also examined the case of $N=2$ with higher dimensional inputs, specifically $d = 3$ and $d = 4$. The parties share an entangled state $\rho_{AB} \in \mathbb{C}^d \otimes \mathbb{C}^d$, Alice's classical message can take on $c \in [d^2]$ distinct values. We numerically lower bound the average fidelity of the corresponding stochastic teleportation protocols. For dimension $d = 3$, we achieve an average fidelity of $F_{\text{avg}} = 0.735$, while for $d = 4$ the fidelity decreases to $F_{\text{avg}} = 0.684$. This suggests that the performance of universal stochastic teleportation protocols decreases with increasing dimension.

\section{Discussion}

A central question in quantum information theory is to characterise quantum communication resources through the correlations that they can generate between separate parties. The ubiquitous scenario for this is the prepare-and-measure scenario. The main focus of the literature on these concerns correlations between classical variables, i.e.~the information that the receiver aims to access is classical data. In this work, we have developed a framework for prepare-and-measure scenarios in which the information itself is of quantum nature. In these scenarios, the sender holds quantum information and the receiver aims to recover some quantum aspect of it. We have established elementary properties that are relevant to any communication task within this framework. Specifically, we discussed how classical and quantum messages exhibit a simple one-to-one relationship, how various performance metrics can be used to benchmark arbitrary quantum protocols and how these protocols can be optimised by numerically. 

In analogy with standard prepare-and-measure scenarios, our framework with quantum inputs accommodates a variety of specific tasks. We have systematically investigated a natural type of task that we call stochastic teleportation, which can equally well be viewed as natural generalisations of either quantum random access codes or quantum teleportation. Our main results show that (i) post-quantum nonlocality can trivialise communication complexity for quantum information, (ii) random access codes and standard teleportation protocols can be systematically transformed into stochastic teleportation protocols, and (iii) that there exists a universal quantum stochastic teleportation machine which permits arbitrary quantum information to be stochastically recovered with high-fidelity. The simplest instance of the latter machine relies on genuine three-qubit entangled measurements that are related neither to the standard GHZ-measurements or the W-basis measurements \cite{Pimpel2023}. This suggests that more complex forms of entangled measurements play a  key role in shaping the understanding of prepare-and-measure scenarios based on quantum information \cite{Pauwels2025}.

Our work leaves several open questions. A selection of these are the following. (1) If Alice's inputs each consist of half of an entangled state, then Alice and Bob will at the end of stochastic teleportation end up in a multipartite entangled state. The entanglement structure of this state must be limited by entanglement monogamy. Understand these restrictions and their implications for stochastic teleportation remains open. (2) Our focus has been on scenarios in which every round of the protocol counts towards the final performance. A relevant further direction is to consider stochastic teleportation protocols in a probabilistic setting, i.e.~when the protocol only succeeds with a non-unit probability. What success probabilities are required to accurately approximate a perfect fidelity in the teleportation? (3) Is every bipartite entangled state a resource in the prepare-and-measure scenario with quantum inputs? Note that the answer is positive in the less demanding situation in which the receiver is a trusted quantum device \cite{Cavalcanti2017}. (4) A central open problem is to develop general methods for bounding the set of quantum correlations that are admissible in prepare-and-measure scenarios with quantum inputs. A promising pathway is to identify dedicated methods based on semidefinite programming relaxations \cite{Tavakoli2024}.

\begin{acknowledgments}
We thank Stefano Pironio for inspiring discussions. This work is supported by the Swedish Research Council under Contract No.~2023-03498, the Knut and Alice Wallenberg Foundation through the Wallenberg Center for Quantum Technology (WACQT) and NCCR-SwissMAP of the Swiss National Science Foundation.
\end{acknowledgments}

\bibliography{references_stochteleport}

\begin{thebibliography}{55}%
\makeatletter
\providecommand \@ifxundefined [1]{%
 \@ifx{#1\undefined}
}%
\providecommand \@ifnum [1]{%
 \ifnum #1\expandafter \@firstoftwo
 \else \expandafter \@secondoftwo
 \fi
}%
\providecommand \@ifx [1]{%
 \ifx #1\expandafter \@firstoftwo
 \else \expandafter \@secondoftwo
 \fi
}%
\providecommand \natexlab [1]{#1}%
\providecommand \enquote  [1]{``#1''}%
\providecommand \bibnamefont  [1]{#1}%
\providecommand \bibfnamefont [1]{#1}%
\providecommand \citenamefont [1]{#1}%
\providecommand \href@noop [0]{\@secondoftwo}%
\providecommand \href [0]{\begingroup \@sanitize@url \@href}%
\providecommand \@href[1]{\@@startlink{#1}\@@href}%
\providecommand \@@href[1]{\endgroup#1\@@endlink}%
\providecommand \@sanitize@url [0]{\catcode `\\12\catcode `\$12\catcode
  `\&12\catcode `\#12\catcode `\^12\catcode `\_12\catcode `\%12\relax}%
\providecommand \@@startlink[1]{}%
\providecommand \@@endlink[0]{}%
\providecommand \url  [0]{\begingroup\@sanitize@url \@url }%
\providecommand \@url [1]{\endgroup\@href {#1}{\urlprefix }}%
\providecommand \urlprefix  [0]{URL }%
\providecommand \Eprint [0]{\href }%
\providecommand \doibase [0]{https://doi.org/}%
\providecommand \selectlanguage [0]{\@gobble}%
\providecommand \bibinfo  [0]{\@secondoftwo}%
\providecommand \bibfield  [0]{\@secondoftwo}%
\providecommand \translation [1]{[#1]}%
\providecommand \BibitemOpen [0]{}%
\providecommand \bibitemStop [0]{}%
\providecommand \bibitemNoStop [0]{.\EOS\space}%
\providecommand \EOS [0]{\spacefactor3000\relax}%
\providecommand \BibitemShut  [1]{\csname bibitem#1\endcsname}%
\let\auto@bib@innerbib\@empty
\bibitem [{\citenamefont {Ambainis}\ \emph {et~al.}(1999)\citenamefont
  {Ambainis}, \citenamefont {Nayak}, \citenamefont {Ta-Shma},\ and\
  \citenamefont {Vazirani}}]{Ambainis1999}%
  \BibitemOpen
  \bibfield  {author} {\bibinfo {author} {\bibfnamefont {A.}~\bibnamefont
  {Ambainis}}, \bibinfo {author} {\bibfnamefont {A.}~\bibnamefont {Nayak}},
  \bibinfo {author} {\bibfnamefont {A.}~\bibnamefont {Ta-Shma}},\ and\ \bibinfo
  {author} {\bibfnamefont {U.}~\bibnamefont {Vazirani}},\ }\bibfield  {title}
  {\bibinfo {title} {Dense quantum coding and a lower bound for 1-way quantum
  automata},\ }in\ \href {https://doi.org/10.1145/301250.301347} {\emph
  {\bibinfo {booktitle} {Proceedings of the Thirty-First Annual ACM Symposium
  on Theory of Computing}}},\ \bibinfo {series and number} {STOC '99}\
  (\bibinfo  {publisher} {Association for Computing Machinery},\ \bibinfo
  {address} {New York, NY, USA},\ \bibinfo {year} {1999})\ p.\ \bibinfo {pages}
  {376–383}\BibitemShut {NoStop}%
\bibitem [{\citenamefont {Nayak}(1999)}]{Nayak1999}%
  \BibitemOpen
  \bibfield  {author} {\bibinfo {author} {\bibfnamefont {A.}~\bibnamefont
  {Nayak}},\ }\bibfield  {title} {\bibinfo {title} {Optimal lower bounds for
  quantum automata and random access codes},\ }in\ \href
  {https://doi.org/10.1109/SFFCS.1999.814608} {\emph {\bibinfo {booktitle}
  {40th Annual Symposium on Foundations of Computer Science (Cat.
  No.99CB37039)}}}\ (\bibinfo {year} {1999})\ pp.\ \bibinfo {pages}
  {369--376}\BibitemShut {NoStop}%
\bibitem [{\citenamefont {Tavakoli}\ \emph {et~al.}(2015)\citenamefont
  {Tavakoli}, \citenamefont {Hameedi}, \citenamefont {Marques},\ and\
  \citenamefont {Bourennane}}]{Tavakoli2015}%
  \BibitemOpen
  \bibfield  {author} {\bibinfo {author} {\bibfnamefont {A.}~\bibnamefont
  {Tavakoli}}, \bibinfo {author} {\bibfnamefont {A.}~\bibnamefont {Hameedi}},
  \bibinfo {author} {\bibfnamefont {B.}~\bibnamefont {Marques}},\ and\ \bibinfo
  {author} {\bibfnamefont {M.}~\bibnamefont {Bourennane}},\ }\bibfield  {title}
  {\bibinfo {title} {Quantum random access codes using single $d$-level
  systems},\ }\href {https://doi.org/10.1103/PhysRevLett.114.170502} {\bibfield
   {journal} {\bibinfo  {journal} {Phys. Rev. Lett.}\ }\textbf {\bibinfo
  {volume} {114}},\ \bibinfo {pages} {170502} (\bibinfo {year}
  {2015})}\BibitemShut {NoStop}%
\bibitem [{\citenamefont {Brunner}\ \emph {et~al.}(2013)\citenamefont
  {Brunner}, \citenamefont {Navascu\'es},\ and\ \citenamefont
  {V\'ertesi}}]{Brunner2013}%
  \BibitemOpen
  \bibfield  {author} {\bibinfo {author} {\bibfnamefont {N.}~\bibnamefont
  {Brunner}}, \bibinfo {author} {\bibfnamefont {M.}~\bibnamefont
  {Navascu\'es}},\ and\ \bibinfo {author} {\bibfnamefont {T.}~\bibnamefont
  {V\'ertesi}},\ }\bibfield  {title} {\bibinfo {title} {Dimension witnesses and
  quantum state discrimination},\ }\href
  {https://doi.org/10.1103/PhysRevLett.110.150501} {\bibfield  {journal}
  {\bibinfo  {journal} {Phys. Rev. Lett.}\ }\textbf {\bibinfo {volume} {110}},\
  \bibinfo {pages} {150501} (\bibinfo {year} {2013})}\BibitemShut {NoStop}%
\bibitem [{\citenamefont {Aguilar}\ \emph {et~al.}(2018)\citenamefont
  {Aguilar}, \citenamefont {Farkas}, \citenamefont {Mart\'{\i}nez},
  \citenamefont {Alvarado}, \citenamefont {Cari\~ne}, \citenamefont {Xavier},
  \citenamefont {Barra}, \citenamefont {Ca\~nas}, \citenamefont
  {Paw\l{}owski},\ and\ \citenamefont {Lima}}]{Aguilar2018}%
  \BibitemOpen
  \bibfield  {author} {\bibinfo {author} {\bibfnamefont {E.~A.}\ \bibnamefont
  {Aguilar}}, \bibinfo {author} {\bibfnamefont {M.}~\bibnamefont {Farkas}},
  \bibinfo {author} {\bibfnamefont {D.}~\bibnamefont {Mart\'{\i}nez}}, \bibinfo
  {author} {\bibfnamefont {M.}~\bibnamefont {Alvarado}}, \bibinfo {author}
  {\bibfnamefont {J.}~\bibnamefont {Cari\~ne}}, \bibinfo {author}
  {\bibfnamefont {G.~B.}\ \bibnamefont {Xavier}}, \bibinfo {author}
  {\bibfnamefont {J.~F.}\ \bibnamefont {Barra}}, \bibinfo {author}
  {\bibfnamefont {G.}~\bibnamefont {Ca\~nas}}, \bibinfo {author} {\bibfnamefont
  {M.}~\bibnamefont {Paw\l{}owski}},\ and\ \bibinfo {author} {\bibfnamefont
  {G.}~\bibnamefont {Lima}},\ }\bibfield  {title} {\bibinfo {title} {Certifying
  an irreducible 1024-dimensional photonic state using refined dimension
  witnesses},\ }\href {https://doi.org/10.1103/PhysRevLett.120.230503}
  {\bibfield  {journal} {\bibinfo  {journal} {Phys. Rev. Lett.}\ }\textbf
  {\bibinfo {volume} {120}},\ \bibinfo {pages} {230503} (\bibinfo {year}
  {2018})}\BibitemShut {NoStop}%
\bibitem [{\citenamefont {Giordani}\ \emph {et~al.}(2023)\citenamefont
  {Giordani}, \citenamefont {Wagner}, \citenamefont {Esposito}, \citenamefont
  {Camillini}, \citenamefont {Hoch}, \citenamefont {Carvacho}, \citenamefont
  {Pentangelo}, \citenamefont {Ceccarelli}, \citenamefont {Piacentini},
  \citenamefont {Crespi}, \citenamefont {Spagnolo}, \citenamefont {Osellame},
  \citenamefont {Galvão},\ and\ \citenamefont {Sciarrino}}]{Giordani2023}%
  \BibitemOpen
  \bibfield  {author} {\bibinfo {author} {\bibfnamefont {T.}~\bibnamefont
  {Giordani}}, \bibinfo {author} {\bibfnamefont {R.}~\bibnamefont {Wagner}},
  \bibinfo {author} {\bibfnamefont {C.}~\bibnamefont {Esposito}}, \bibinfo
  {author} {\bibfnamefont {A.}~\bibnamefont {Camillini}}, \bibinfo {author}
  {\bibfnamefont {F.}~\bibnamefont {Hoch}}, \bibinfo {author} {\bibfnamefont
  {G.}~\bibnamefont {Carvacho}}, \bibinfo {author} {\bibfnamefont
  {C.}~\bibnamefont {Pentangelo}}, \bibinfo {author} {\bibfnamefont
  {F.}~\bibnamefont {Ceccarelli}}, \bibinfo {author} {\bibfnamefont
  {S.}~\bibnamefont {Piacentini}}, \bibinfo {author} {\bibfnamefont
  {A.}~\bibnamefont {Crespi}}, \bibinfo {author} {\bibfnamefont
  {N.}~\bibnamefont {Spagnolo}}, \bibinfo {author} {\bibfnamefont
  {R.}~\bibnamefont {Osellame}}, \bibinfo {author} {\bibfnamefont {E.~F.}\
  \bibnamefont {Galvão}},\ and\ \bibinfo {author} {\bibfnamefont
  {F.}~\bibnamefont {Sciarrino}},\ }\bibfield  {title} {\bibinfo {title}
  {Experimental certification of contextuality, coherence, and dimension in a
  programmable universal photonic processor},\ }\href
  {https://doi.org/10.1126/sciadv.adj4249} {\bibfield  {journal} {\bibinfo
  {journal} {Science Advances}\ }\textbf {\bibinfo {volume} {9}},\ \bibinfo
  {pages} {eadj4249} (\bibinfo {year} {2023})},\ \Eprint
  {https://arxiv.org/abs/https://www.science.org/doi/pdf/10.1126/sciadv.adj4249}
  {https://www.science.org/doi/pdf/10.1126/sciadv.adj4249} \BibitemShut
  {NoStop}%
\bibitem [{\citenamefont {Bernal}\ \emph {et~al.}(2024)\citenamefont {Bernal},
  \citenamefont {Cobucci}, \citenamefont {Renner},\ and\ \citenamefont
  {Tavakoli}}]{Bernal2024}%
  \BibitemOpen
  \bibfield  {author} {\bibinfo {author} {\bibfnamefont {A.}~\bibnamefont
  {Bernal}}, \bibinfo {author} {\bibfnamefont {G.}~\bibnamefont {Cobucci}},
  \bibinfo {author} {\bibfnamefont {M.~J.}\ \bibnamefont {Renner}},\ and\
  \bibinfo {author} {\bibfnamefont {A.}~\bibnamefont {Tavakoli}},\ }\bibfield
  {title} {\bibinfo {title} {Absolute dimensionality of quantum ensembles},\
  }\href {https://doi.org/10.1103/PhysRevLett.133.240203} {\bibfield  {journal}
  {\bibinfo  {journal} {Phys. Rev. Lett.}\ }\textbf {\bibinfo {volume} {133}},\
  \bibinfo {pages} {240203} (\bibinfo {year} {2024})}\BibitemShut {NoStop}%
\bibitem [{\citenamefont {H\aa{}kansson}\ \emph {et~al.}(2025)\citenamefont
  {H\aa{}kansson}, \citenamefont {Piveteau}, \citenamefont {Seguinard},
  \citenamefont {Muhammad}, \citenamefont {Bourennane}, \citenamefont
  {G\"uhne},\ and\ \citenamefont {Pl\'avala}}]{Hakanson2025}%
  \BibitemOpen
  \bibfield  {author} {\bibinfo {author} {\bibfnamefont {E.}~\bibnamefont
  {H\aa{}kansson}}, \bibinfo {author} {\bibfnamefont {A.}~\bibnamefont
  {Piveteau}}, \bibinfo {author} {\bibfnamefont {A.}~\bibnamefont {Seguinard}},
  \bibinfo {author} {\bibfnamefont {S.}~\bibnamefont {Muhammad}}, \bibinfo
  {author} {\bibfnamefont {M.}~\bibnamefont {Bourennane}}, \bibinfo {author}
  {\bibfnamefont {O.}~\bibnamefont {G\"uhne}},\ and\ \bibinfo {author}
  {\bibfnamefont {M.}~\bibnamefont {Pl\'avala}},\ }\bibfield  {title} {\bibinfo
  {title} {Experimental implementation of dimension-dependent contextuality
  inequality},\ }\href {https://doi.org/10.1103/PhysRevLett.134.200202}
  {\bibfield  {journal} {\bibinfo  {journal} {Phys. Rev. Lett.}\ }\textbf
  {\bibinfo {volume} {134}},\ \bibinfo {pages} {200202} (\bibinfo {year}
  {2025})}\BibitemShut {NoStop}%
\bibitem [{\citenamefont {Brukner}\ \emph {et~al.}(2004)\citenamefont
  {Brukner}, \citenamefont {\ifmmode~\dot{Z}\else \.{Z}\fi{}ukowski},
  \citenamefont {Pan},\ and\ \citenamefont {Zeilinger}}]{Brukner2004}%
  \BibitemOpen
  \bibfield  {author} {\bibinfo {author} {\bibfnamefont {i.~c.~v.}\
  \bibnamefont {Brukner}}, \bibinfo {author} {\bibfnamefont {M.}~\bibnamefont
  {\ifmmode~\dot{Z}\else \.{Z}\fi{}ukowski}}, \bibinfo {author} {\bibfnamefont
  {J.-W.}\ \bibnamefont {Pan}},\ and\ \bibinfo {author} {\bibfnamefont
  {A.}~\bibnamefont {Zeilinger}},\ }\bibfield  {title} {\bibinfo {title}
  {Bell's inequalities and quantum communication complexity},\ }\href
  {https://doi.org/10.1103/PhysRevLett.92.127901} {\bibfield  {journal}
  {\bibinfo  {journal} {Phys. Rev. Lett.}\ }\textbf {\bibinfo {volume} {92}},\
  \bibinfo {pages} {127901} (\bibinfo {year} {2004})}\BibitemShut {NoStop}%
\bibitem [{\citenamefont {Tavakoli}\ \emph
  {et~al.}(2020{\natexlab{a}})\citenamefont {Tavakoli}, \citenamefont
  {{\.{Z}}ukowski},\ and\ \citenamefont {Brukner}}]{Tavakoli2020}%
  \BibitemOpen
  \bibfield  {author} {\bibinfo {author} {\bibfnamefont {A.}~\bibnamefont
  {Tavakoli}}, \bibinfo {author} {\bibfnamefont {M.}~\bibnamefont
  {{\.{Z}}ukowski}},\ and\ \bibinfo {author} {\bibfnamefont
  {{\v{C}}.}~\bibnamefont {Brukner}},\ }\bibfield  {title} {\bibinfo {title}
  {Does violation of a {B}ell inequality always imply quantum advantage in a
  communication complexity problem?},\ }\href
  {https://doi.org/10.22331/q-2020-09-07-316} {\bibfield  {journal} {\bibinfo
  {journal} {{Quantum}}\ }\textbf {\bibinfo {volume} {4}},\ \bibinfo {pages}
  {316} (\bibinfo {year} {2020}{\natexlab{a}})}\BibitemShut {NoStop}%
\bibitem [{\citenamefont {Pauwels}\ \emph
  {et~al.}(2022{\natexlab{a}})\citenamefont {Pauwels}, \citenamefont {Pironio},
  \citenamefont {Cruzeiro},\ and\ \citenamefont {Tavakoli}}]{Pauwels2022}%
  \BibitemOpen
  \bibfield  {author} {\bibinfo {author} {\bibfnamefont {J.}~\bibnamefont
  {Pauwels}}, \bibinfo {author} {\bibfnamefont {S.}~\bibnamefont {Pironio}},
  \bibinfo {author} {\bibfnamefont {E.~Z.}\ \bibnamefont {Cruzeiro}},\ and\
  \bibinfo {author} {\bibfnamefont {A.}~\bibnamefont {Tavakoli}},\ }\bibfield
  {title} {\bibinfo {title} {Adaptive advantage in entanglement-assisted
  communications},\ }\href {https://doi.org/10.1103/PhysRevLett.129.120504}
  {\bibfield  {journal} {\bibinfo  {journal} {Phys. Rev. Lett.}\ }\textbf
  {\bibinfo {volume} {129}},\ \bibinfo {pages} {120504} (\bibinfo {year}
  {2022}{\natexlab{a}})}\BibitemShut {NoStop}%
\bibitem [{\citenamefont {Paw\l{}owski}\ and\ \citenamefont
  {\ifmmode~\dot{Z}\else \.{Z}\fi{}ukowski}(2010)}]{Pawlowski2010}%
  \BibitemOpen
  \bibfield  {author} {\bibinfo {author} {\bibfnamefont {M.}~\bibnamefont
  {Paw\l{}owski}}\ and\ \bibinfo {author} {\bibfnamefont {M.}~\bibnamefont
  {\ifmmode~\dot{Z}\else \.{Z}\fi{}ukowski}},\ }\bibfield  {title} {\bibinfo
  {title} {Entanglement-assisted random access codes},\ }\href
  {https://doi.org/10.1103/PhysRevA.81.042326} {\bibfield  {journal} {\bibinfo
  {journal} {Phys. Rev. A}\ }\textbf {\bibinfo {volume} {81}},\ \bibinfo
  {pages} {042326} (\bibinfo {year} {2010})}\BibitemShut {NoStop}%
\bibitem [{\citenamefont {Mart\'{\i}nez}\ \emph {et~al.}(2018)\citenamefont
  {Mart\'{\i}nez}, \citenamefont {Tavakoli}, \citenamefont {Casanova},
  \citenamefont {Ca\~nas}, \citenamefont {Marques},\ and\ \citenamefont
  {Lima}}]{Martinez2018}%
  \BibitemOpen
  \bibfield  {author} {\bibinfo {author} {\bibfnamefont {D.}~\bibnamefont
  {Mart\'{\i}nez}}, \bibinfo {author} {\bibfnamefont {A.}~\bibnamefont
  {Tavakoli}}, \bibinfo {author} {\bibfnamefont {M.}~\bibnamefont {Casanova}},
  \bibinfo {author} {\bibfnamefont {G.}~\bibnamefont {Ca\~nas}}, \bibinfo
  {author} {\bibfnamefont {B.}~\bibnamefont {Marques}},\ and\ \bibinfo {author}
  {\bibfnamefont {G.}~\bibnamefont {Lima}},\ }\bibfield  {title} {\bibinfo
  {title} {High-dimensional quantum communication complexity beyond strategies
  based on bell's theorem},\ }\href
  {https://doi.org/10.1103/PhysRevLett.121.150504} {\bibfield  {journal}
  {\bibinfo  {journal} {Phys. Rev. Lett.}\ }\textbf {\bibinfo {volume} {121}},\
  \bibinfo {pages} {150504} (\bibinfo {year} {2018})}\BibitemShut {NoStop}%
\bibitem [{\citenamefont {Tavakoli}\ \emph {et~al.}(2021)\citenamefont
  {Tavakoli}, \citenamefont {Pauwels}, \citenamefont {Woodhead},\ and\
  \citenamefont {Pironio}}]{Tavakoli2021}%
  \BibitemOpen
  \bibfield  {author} {\bibinfo {author} {\bibfnamefont {A.}~\bibnamefont
  {Tavakoli}}, \bibinfo {author} {\bibfnamefont {J.}~\bibnamefont {Pauwels}},
  \bibinfo {author} {\bibfnamefont {E.}~\bibnamefont {Woodhead}},\ and\
  \bibinfo {author} {\bibfnamefont {S.}~\bibnamefont {Pironio}},\ }\bibfield
  {title} {\bibinfo {title} {Correlations in entanglement-assisted
  prepare-and-measure scenarios},\ }\href
  {https://doi.org/10.1103/PRXQuantum.2.040357} {\bibfield  {journal} {\bibinfo
   {journal} {PRX Quantum}\ }\textbf {\bibinfo {volume} {2}},\ \bibinfo {pages}
  {040357} (\bibinfo {year} {2021})}\BibitemShut {NoStop}%
\bibitem [{\citenamefont {Pauwels}\ \emph
  {et~al.}(2022{\natexlab{b}})\citenamefont {Pauwels}, \citenamefont
  {Tavakoli}, \citenamefont {Woodhead},\ and\ \citenamefont
  {Pironio}}]{Pauwels2022b}%
  \BibitemOpen
  \bibfield  {author} {\bibinfo {author} {\bibfnamefont {J.}~\bibnamefont
  {Pauwels}}, \bibinfo {author} {\bibfnamefont {A.}~\bibnamefont {Tavakoli}},
  \bibinfo {author} {\bibfnamefont {E.}~\bibnamefont {Woodhead}},\ and\
  \bibinfo {author} {\bibfnamefont {S.}~\bibnamefont {Pironio}},\ }\bibfield
  {title} {\bibinfo {title} {Entanglement in prepare-and-measure scenarios:
  many questions, a few answers},\ }\href
  {https://doi.org/10.1088/1367-2630/ac724a} {\bibfield  {journal} {\bibinfo
  {journal} {New Journal of Physics}\ }\textbf {\bibinfo {volume} {24}},\
  \bibinfo {pages} {063015} (\bibinfo {year} {2022}{\natexlab{b}})}\BibitemShut
  {NoStop}%
\bibitem [{\citenamefont {Bennett}\ and\ \citenamefont
  {Wiesner}(1992)}]{Bennett1992}%
  \BibitemOpen
  \bibfield  {author} {\bibinfo {author} {\bibfnamefont {C.~H.}\ \bibnamefont
  {Bennett}}\ and\ \bibinfo {author} {\bibfnamefont {S.~J.}\ \bibnamefont
  {Wiesner}},\ }\bibfield  {title} {\bibinfo {title} {Communication via one-
  and two-particle operators on einstein-podolsky-rosen states},\ }\href
  {https://doi.org/10.1103/PhysRevLett.69.2881} {\bibfield  {journal} {\bibinfo
   {journal} {Phys. Rev. Lett.}\ }\textbf {\bibinfo {volume} {69}},\ \bibinfo
  {pages} {2881} (\bibinfo {year} {1992})}\BibitemShut {NoStop}%
\bibitem [{\citenamefont {Piveteau}\ \emph {et~al.}(2022)\citenamefont
  {Piveteau}, \citenamefont {Pauwels}, \citenamefont {H{\aa}kansson},
  \citenamefont {Muhammad}, \citenamefont {Bourennane},\ and\ \citenamefont
  {Tavakoli}}]{Piveteau2022}%
  \BibitemOpen
  \bibfield  {author} {\bibinfo {author} {\bibfnamefont {A.}~\bibnamefont
  {Piveteau}}, \bibinfo {author} {\bibfnamefont {J.}~\bibnamefont {Pauwels}},
  \bibinfo {author} {\bibfnamefont {E.}~\bibnamefont {H{\aa}kansson}}, \bibinfo
  {author} {\bibfnamefont {S.}~\bibnamefont {Muhammad}}, \bibinfo {author}
  {\bibfnamefont {M.}~\bibnamefont {Bourennane}},\ and\ \bibinfo {author}
  {\bibfnamefont {A.}~\bibnamefont {Tavakoli}},\ }\bibfield  {title} {\bibinfo
  {title} {Entanglement-assisted quantum communication with simple
  measurements},\ }\href {https://doi.org/10.1038/s41467-022-33922-5}
  {\bibfield  {journal} {\bibinfo  {journal} {Nature Communications}\ }\textbf
  {\bibinfo {volume} {13}},\ \bibinfo {pages} {7878} (\bibinfo {year}
  {2022})}\BibitemShut {NoStop}%
\bibitem [{\citenamefont {Bakhshinezhad}\ \emph {et~al.}(2024)\citenamefont
  {Bakhshinezhad}, \citenamefont {Mehboudi}, \citenamefont {Carceller},\ and\
  \citenamefont {Tavakoli}}]{Bakhshinezhad2024}%
  \BibitemOpen
  \bibfield  {author} {\bibinfo {author} {\bibfnamefont {P.}~\bibnamefont
  {Bakhshinezhad}}, \bibinfo {author} {\bibfnamefont {M.}~\bibnamefont
  {Mehboudi}}, \bibinfo {author} {\bibfnamefont {C.~R.~i.}\ \bibnamefont
  {Carceller}},\ and\ \bibinfo {author} {\bibfnamefont {A.}~\bibnamefont
  {Tavakoli}},\ }\bibfield  {title} {\bibinfo {title} {Scalable entanglement
  certification via quantum communication},\ }\href
  {https://doi.org/10.1103/PRXQuantum.5.020319} {\bibfield  {journal} {\bibinfo
   {journal} {PRX Quantum}\ }\textbf {\bibinfo {volume} {5}},\ \bibinfo {pages}
  {020319} (\bibinfo {year} {2024})}\BibitemShut {NoStop}%
\bibitem [{\citenamefont {Zhang}\ \emph {et~al.}(2025)\citenamefont {Zhang},
  \citenamefont {Miao}, \citenamefont {Hu}, \citenamefont {Pauwels},
  \citenamefont {Guo}, \citenamefont {Li}, \citenamefont {Guo}, \citenamefont
  {Tavakoli},\ and\ \citenamefont {Liu}}]{Zhang2025}%
  \BibitemOpen
  \bibfield  {author} {\bibinfo {author} {\bibfnamefont {C.}~\bibnamefont
  {Zhang}}, \bibinfo {author} {\bibfnamefont {J.-L.}\ \bibnamefont {Miao}},
  \bibinfo {author} {\bibfnamefont {X.-M.}\ \bibnamefont {Hu}}, \bibinfo
  {author} {\bibfnamefont {J.}~\bibnamefont {Pauwels}}, \bibinfo {author}
  {\bibfnamefont {Y.}~\bibnamefont {Guo}}, \bibinfo {author} {\bibfnamefont
  {C.-F.}\ \bibnamefont {Li}}, \bibinfo {author} {\bibfnamefont {G.-C.}\
  \bibnamefont {Guo}}, \bibinfo {author} {\bibfnamefont {A.}~\bibnamefont
  {Tavakoli}},\ and\ \bibinfo {author} {\bibfnamefont {B.-H.}\ \bibnamefont
  {Liu}},\ }\bibfield  {title} {\bibinfo {title} {Quantum stochastic
  communication via high-dimensional entanglement},\ }\href
  {https://doi.org/10.1103/rq78-1qbh} {\bibfield  {journal} {\bibinfo
  {journal} {Phys. Rev. Lett.}\ }\textbf {\bibinfo {volume} {135}},\ \bibinfo
  {pages} {120802} (\bibinfo {year} {2025})}\BibitemShut {NoStop}%
\bibitem [{\citenamefont {Bennett}\ \emph {et~al.}(1993)\citenamefont
  {Bennett}, \citenamefont {Brassard}, \citenamefont {Cr\'epeau}, \citenamefont
  {Jozsa}, \citenamefont {Peres},\ and\ \citenamefont
  {Wootters}}]{Bennett1993}%
  \BibitemOpen
  \bibfield  {author} {\bibinfo {author} {\bibfnamefont {C.~H.}\ \bibnamefont
  {Bennett}}, \bibinfo {author} {\bibfnamefont {G.}~\bibnamefont {Brassard}},
  \bibinfo {author} {\bibfnamefont {C.}~\bibnamefont {Cr\'epeau}}, \bibinfo
  {author} {\bibfnamefont {R.}~\bibnamefont {Jozsa}}, \bibinfo {author}
  {\bibfnamefont {A.}~\bibnamefont {Peres}},\ and\ \bibinfo {author}
  {\bibfnamefont {W.~K.}\ \bibnamefont {Wootters}},\ }\bibfield  {title}
  {\bibinfo {title} {Teleporting an unknown quantum state via dual classical
  and einstein-podolsky-rosen channels},\ }\href
  {https://doi.org/10.1103/PhysRevLett.70.1895} {\bibfield  {journal} {\bibinfo
   {journal} {Phys. Rev. Lett.}\ }\textbf {\bibinfo {volume} {70}},\ \bibinfo
  {pages} {1895} (\bibinfo {year} {1993})}\BibitemShut {NoStop}%
\bibitem [{\citenamefont {Pirandola}\ \emph {et~al.}(2015)\citenamefont
  {Pirandola}, \citenamefont {Eisert}, \citenamefont {Weedbrook}, \citenamefont
  {Furusawa},\ and\ \citenamefont {Braunstein}}]{Pirandola2015}%
  \BibitemOpen
  \bibfield  {author} {\bibinfo {author} {\bibfnamefont {S.}~\bibnamefont
  {Pirandola}}, \bibinfo {author} {\bibfnamefont {J.}~\bibnamefont {Eisert}},
  \bibinfo {author} {\bibfnamefont {C.}~\bibnamefont {Weedbrook}}, \bibinfo
  {author} {\bibfnamefont {A.}~\bibnamefont {Furusawa}},\ and\ \bibinfo
  {author} {\bibfnamefont {S.~L.}\ \bibnamefont {Braunstein}},\ }\bibfield
  {title} {\bibinfo {title} {Advances in quantum teleportation},\ }\href
  {https://doi.org/10.1038/nphoton.2015.154} {\bibfield  {journal} {\bibinfo
  {journal} {Nature Photonics}\ }\textbf {\bibinfo {volume} {9}},\ \bibinfo
  {pages} {641} (\bibinfo {year} {2015})}\BibitemShut {NoStop}%
\bibitem [{\citenamefont {Hu}\ \emph {et~al.}(2023)\citenamefont {Hu},
  \citenamefont {Guo}, \citenamefont {Liu}, \citenamefont {Li},\ and\
  \citenamefont {Guo}}]{Hu2023}%
  \BibitemOpen
  \bibfield  {author} {\bibinfo {author} {\bibfnamefont {X.-M.}\ \bibnamefont
  {Hu}}, \bibinfo {author} {\bibfnamefont {Y.}~\bibnamefont {Guo}}, \bibinfo
  {author} {\bibfnamefont {B.-H.}\ \bibnamefont {Liu}}, \bibinfo {author}
  {\bibfnamefont {C.-F.}\ \bibnamefont {Li}},\ and\ \bibinfo {author}
  {\bibfnamefont {G.-C.}\ \bibnamefont {Guo}},\ }\bibfield  {title} {\bibinfo
  {title} {Progress in quantum teleportation},\ }\href
  {https://doi.org/10.1038/s42254-023-00588-x} {\bibfield  {journal} {\bibinfo
  {journal} {Nature Reviews Physics}\ }\textbf {\bibinfo {volume} {5}},\
  \bibinfo {pages} {339} (\bibinfo {year} {2023})}\BibitemShut {NoStop}%
\bibitem [{\citenamefont {Pital\'ua-Garc\'{\i}a}(2013)}]{Pitalua2013}%
  \BibitemOpen
  \bibfield  {author} {\bibinfo {author} {\bibfnamefont {D.}~\bibnamefont
  {Pital\'ua-Garc\'{\i}a}},\ }\bibfield  {title} {\bibinfo {title} {Quantum
  information causality},\ }\href
  {https://doi.org/10.1103/PhysRevLett.110.210402} {\bibfield  {journal}
  {\bibinfo  {journal} {Phys. Rev. Lett.}\ }\textbf {\bibinfo {volume} {110}},\
  \bibinfo {pages} {210402} (\bibinfo {year} {2013})}\BibitemShut {NoStop}%
\bibitem [{\citenamefont {Sakharwade}\ \emph {et~al.}(2023)\citenamefont
  {Sakharwade}, \citenamefont {Studziński}, \citenamefont {Eckstein},\ and\
  \citenamefont {Horodecki}}]{Sakharwade2023}%
  \BibitemOpen
  \bibfield  {author} {\bibinfo {author} {\bibfnamefont {N.}~\bibnamefont
  {Sakharwade}}, \bibinfo {author} {\bibfnamefont {M.}~\bibnamefont
  {Studziński}}, \bibinfo {author} {\bibfnamefont {M.}~\bibnamefont
  {Eckstein}},\ and\ \bibinfo {author} {\bibfnamefont {P.}~\bibnamefont
  {Horodecki}},\ }\bibfield  {title} {\bibinfo {title} {Two instances of random
  access code in the quantum regime},\ }\href
  {https://doi.org/10.1088/1367-2630/acd716} {\bibfield  {journal} {\bibinfo
  {journal} {New Journal of Physics}\ }\textbf {\bibinfo {volume} {25}},\
  \bibinfo {pages} {053038} (\bibinfo {year} {2023})}\BibitemShut {NoStop}%
\bibitem [{\citenamefont {Grudka}\ \emph {et~al.}(2015)\citenamefont {Grudka},
  \citenamefont {Horodecki}, \citenamefont {Horodecki},\ and\ \citenamefont
  {W\'ojcik}}]{Grudka2015}%
  \BibitemOpen
  \bibfield  {author} {\bibinfo {author} {\bibfnamefont {A.}~\bibnamefont
  {Grudka}}, \bibinfo {author} {\bibfnamefont {M.}~\bibnamefont {Horodecki}},
  \bibinfo {author} {\bibfnamefont {R.}~\bibnamefont {Horodecki}},\ and\
  \bibinfo {author} {\bibfnamefont {A.}~\bibnamefont {W\'ojcik}},\ }\bibfield
  {title} {\bibinfo {title} {Nonsignaling quantum random access-code boxes},\
  }\href {https://doi.org/10.1103/PhysRevA.92.052312} {\bibfield  {journal}
  {\bibinfo  {journal} {Phys. Rev. A}\ }\textbf {\bibinfo {volume} {92}},\
  \bibinfo {pages} {052312} (\bibinfo {year} {2015})}\BibitemShut {NoStop}%
\bibitem [{\citenamefont {Popescu}\ and\ \citenamefont
  {Rohrlich}(1994)}]{Popescu1994}%
  \BibitemOpen
  \bibfield  {author} {\bibinfo {author} {\bibfnamefont {S.}~\bibnamefont
  {Popescu}}\ and\ \bibinfo {author} {\bibfnamefont {D.}~\bibnamefont
  {Rohrlich}},\ }\bibfield  {title} {\bibinfo {title} {Quantum nonlocality as
  an axiom},\ }\href {https://doi.org/10.1007/BF02058098} {\bibfield  {journal}
  {\bibinfo  {journal} {Foundations of Physics}\ }\textbf {\bibinfo {volume}
  {24}},\ \bibinfo {pages} {379} (\bibinfo {year} {1994})}\BibitemShut
  {NoStop}%
\bibitem [{\citenamefont {van Dam}(2013)}]{vanDam2013}%
  \BibitemOpen
  \bibfield  {author} {\bibinfo {author} {\bibfnamefont {W.}~\bibnamefont {van
  Dam}},\ }\bibfield  {title} {\bibinfo {title} {Implausible consequences of
  superstrong nonlocality},\ }\href {https://doi.org/10.1007/s11047-012-9353-6}
  {\bibfield  {journal} {\bibinfo  {journal} {Natural Computing}\ }\textbf
  {\bibinfo {volume} {12}},\ \bibinfo {pages} {9} (\bibinfo {year}
  {2013})}\BibitemShut {NoStop}%
\bibitem [{\citenamefont {Vieira}\ \emph {et~al.}(2023)\citenamefont {Vieira},
  \citenamefont {de~Gois}, \citenamefont {Pollyceno},\ and\ \citenamefont
  {Rabelo}}]{Vieira2023}%
  \BibitemOpen
  \bibfield  {author} {\bibinfo {author} {\bibfnamefont {C.}~\bibnamefont
  {Vieira}}, \bibinfo {author} {\bibfnamefont {C.}~\bibnamefont {de~Gois}},
  \bibinfo {author} {\bibfnamefont {L.}~\bibnamefont {Pollyceno}},\ and\
  \bibinfo {author} {\bibfnamefont {R.}~\bibnamefont {Rabelo}},\ }\bibfield
  {title} {\bibinfo {title} {Interplays between classical and quantum
  entanglement-assisted communication scenarios},\ }\href
  {https://doi.org/10.1088/1367-2630/ad0526} {\bibfield  {journal} {\bibinfo
  {journal} {New Journal of Physics}\ }\textbf {\bibinfo {volume} {25}},\
  \bibinfo {pages} {113004} (\bibinfo {year} {2023})}\BibitemShut {NoStop}%
\bibitem [{\citenamefont {Horodecki}\ \emph {et~al.}(1999)\citenamefont
  {Horodecki}, \citenamefont {Horodecki},\ and\ \citenamefont
  {Horodecki}}]{Horodecki1999}%
  \BibitemOpen
  \bibfield  {author} {\bibinfo {author} {\bibfnamefont {M.}~\bibnamefont
  {Horodecki}}, \bibinfo {author} {\bibfnamefont {P.}~\bibnamefont
  {Horodecki}},\ and\ \bibinfo {author} {\bibfnamefont {R.}~\bibnamefont
  {Horodecki}},\ }\bibfield  {title} {\bibinfo {title} {General teleportation
  channel, singlet fraction, and quasidistillation},\ }\href
  {https://doi.org/10.1103/PhysRevA.60.1888} {\bibfield  {journal} {\bibinfo
  {journal} {Phys. Rev. A}\ }\textbf {\bibinfo {volume} {60}},\ \bibinfo
  {pages} {1888} (\bibinfo {year} {1999})}\BibitemShut {NoStop}%
\bibitem [{\citenamefont {Werner}(1998)}]{Werner1998}%
  \BibitemOpen
  \bibfield  {author} {\bibinfo {author} {\bibfnamefont {R.~F.}\ \bibnamefont
  {Werner}},\ }\bibfield  {title} {\bibinfo {title} {Optimal cloning of pure
  states},\ }\href {https://doi.org/10.1103/PhysRevA.58.1827} {\bibfield
  {journal} {\bibinfo  {journal} {Phys. Rev. A}\ }\textbf {\bibinfo {volume}
  {58}},\ \bibinfo {pages} {1827} (\bibinfo {year} {1998})}\BibitemShut
  {NoStop}%
\bibitem [{\citenamefont {Tavakoli}\ \emph {et~al.}(2024)\citenamefont
  {Tavakoli}, \citenamefont {Pozas-Kerstjens}, \citenamefont {Brown},\ and\
  \citenamefont {Ara\'ujo}}]{Tavakoli2024}%
  \BibitemOpen
  \bibfield  {author} {\bibinfo {author} {\bibfnamefont {A.}~\bibnamefont
  {Tavakoli}}, \bibinfo {author} {\bibfnamefont {A.}~\bibnamefont
  {Pozas-Kerstjens}}, \bibinfo {author} {\bibfnamefont {P.}~\bibnamefont
  {Brown}},\ and\ \bibinfo {author} {\bibfnamefont {M.}~\bibnamefont
  {Ara\'ujo}},\ }\bibfield  {title} {\bibinfo {title} {Semidefinite programming
  relaxations for quantum correlations},\ }\href
  {https://doi.org/10.1103/RevModPhys.96.045006} {\bibfield  {journal}
  {\bibinfo  {journal} {Rev. Mod. Phys.}\ }\textbf {\bibinfo {volume} {96}},\
  \bibinfo {pages} {045006} (\bibinfo {year} {2024})}\BibitemShut {NoStop}%
\bibitem [{\citenamefont {Skrzypczyk}\ and\ \citenamefont
  {Cavalcanti}(2023{\natexlab{a}})}]{Skrzypczyk2023}%
  \BibitemOpen
  \bibfield  {author} {\bibinfo {author} {\bibfnamefont {P.}~\bibnamefont
  {Skrzypczyk}}\ and\ \bibinfo {author} {\bibfnamefont {D.}~\bibnamefont
  {Cavalcanti}},\ }\href {https://doi.org/10.1088/978-0-7503-3343-6} {\emph
  {\bibinfo {title} {Semidefinite Programming in Quantum Information
  Science}}}\ (\bibinfo  {publisher} {IOP Publishing},\ \bibinfo {year}
  {2023})\BibitemShut {NoStop}%
\bibitem [{\citenamefont {Skrzypczyk}\ and\ \citenamefont
  {Cavalcanti}(2023{\natexlab{b}})}]{SkrzypczykBook}%
  \BibitemOpen
  \bibfield  {author} {\bibinfo {author} {\bibfnamefont {P.}~\bibnamefont
  {Skrzypczyk}}\ and\ \bibinfo {author} {\bibfnamefont {D.}~\bibnamefont
  {Cavalcanti}},\ }\href {https://doi.org/10.1088/978-0-7503-3343-6} {\emph
  {\bibinfo {title} {Semidefinite Programming in Quantum Information
  Science}}},\ 2053-2563\ (\bibinfo  {publisher} {IOP Publishing},\ \bibinfo
  {year} {2023})\BibitemShut {NoStop}%
\bibitem [{\citenamefont {Renes}\ \emph {et~al.}(2004)\citenamefont {Renes},
  \citenamefont {Blume-Kohout}, \citenamefont {Scott},\ and\ \citenamefont
  {Caves}}]{Renes2004}%
  \BibitemOpen
  \bibfield  {author} {\bibinfo {author} {\bibfnamefont {J.~M.}\ \bibnamefont
  {Renes}}, \bibinfo {author} {\bibfnamefont {R.}~\bibnamefont {Blume-Kohout}},
  \bibinfo {author} {\bibfnamefont {A.~J.}\ \bibnamefont {Scott}},\ and\
  \bibinfo {author} {\bibfnamefont {C.~M.}\ \bibnamefont {Caves}},\ }\bibfield
  {title} {\bibinfo {title} {Symmetric informationally complete quantum
  measurements},\ }\href {https://doi.org/10.1063/1.1737053} {\bibfield
  {journal} {\bibinfo  {journal} {Journal of Mathematical Physics}\ }\textbf
  {\bibinfo {volume} {45}},\ \bibinfo {pages} {2171–2180} (\bibinfo {year}
  {2004})}\BibitemShut {NoStop}%
\bibitem [{\citenamefont {Scott}\ and\ \citenamefont
  {Grassl}(2010)}]{Scott2010}%
  \BibitemOpen
  \bibfield  {author} {\bibinfo {author} {\bibfnamefont {A.~J.}\ \bibnamefont
  {Scott}}\ and\ \bibinfo {author} {\bibfnamefont {M.}~\bibnamefont {Grassl}},\
  }\bibfield  {title} {\bibinfo {title} {Symmetric informationally complete
  positive-operator-valued measures: A new computer study},\ }\href
  {https://doi.org/10.1063/1.3374022} {\bibfield  {journal} {\bibinfo
  {journal} {Journal of Mathematical Physics}\ }\textbf {\bibinfo {volume}
  {51}},\ \bibinfo {pages} {042203} (\bibinfo {year} {2010})}\BibitemShut
  {NoStop}%
\bibitem [{\citenamefont {Paw{\l}owski}\ \emph {et~al.}(2009)\citenamefont
  {Paw{\l}owski}, \citenamefont {Paterek}, \citenamefont {Kaszlikowski},
  \citenamefont {Scarani}, \citenamefont {Winter},\ and\ \citenamefont
  {{\.{Z}}ukowski}}]{Pawlowski2009}%
  \BibitemOpen
  \bibfield  {author} {\bibinfo {author} {\bibfnamefont {M.}~\bibnamefont
  {Paw{\l}owski}}, \bibinfo {author} {\bibfnamefont {T.}~\bibnamefont
  {Paterek}}, \bibinfo {author} {\bibfnamefont {D.}~\bibnamefont
  {Kaszlikowski}}, \bibinfo {author} {\bibfnamefont {V.}~\bibnamefont
  {Scarani}}, \bibinfo {author} {\bibfnamefont {A.}~\bibnamefont {Winter}},\
  and\ \bibinfo {author} {\bibfnamefont {M.}~\bibnamefont {{\.{Z}}ukowski}},\
  }\bibfield  {title} {\bibinfo {title} {Information causality as a physical
  principle},\ }\href {https://doi.org/10.1038/nature08400} {\bibfield
  {journal} {\bibinfo  {journal} {Nature}\ }\textbf {\bibinfo {volume} {461}},\
  \bibinfo {pages} {1101} (\bibinfo {year} {2009})}\BibitemShut {NoStop}%
\bibitem [{\citenamefont {Chailloux}\ \emph {et~al.}(2016)\citenamefont
  {Chailloux}, \citenamefont {Kerenidis}, \citenamefont {Kundu},\ and\
  \citenamefont {Sikora}}]{Chailloux2016}%
  \BibitemOpen
  \bibfield  {author} {\bibinfo {author} {\bibfnamefont {A.}~\bibnamefont
  {Chailloux}}, \bibinfo {author} {\bibfnamefont {I.}~\bibnamefont
  {Kerenidis}}, \bibinfo {author} {\bibfnamefont {S.}~\bibnamefont {Kundu}},\
  and\ \bibinfo {author} {\bibfnamefont {J.}~\bibnamefont {Sikora}},\
  }\bibfield  {title} {\bibinfo {title} {Optimal bounds for parity-oblivious
  random access codes},\ }\href {https://doi.org/10.1088/1367-2630/18/4/045003}
  {\bibfield  {journal} {\bibinfo  {journal} {New Journal of Physics}\ }\textbf
  {\bibinfo {volume} {18}},\ \bibinfo {pages} {045003} (\bibinfo {year}
  {2016})}\BibitemShut {NoStop}%
\bibitem [{\citenamefont {Muhammad}\ \emph {et~al.}(2014)\citenamefont
  {Muhammad}, \citenamefont {Tavakoli}, \citenamefont {Kurant}, \citenamefont
  {Paw\l{}owski}, \citenamefont {\ifmmode~\dot{Z}\else \.{Z}\fi{}ukowski},\
  and\ \citenamefont {Bourennane}}]{Sadiq2014}%
  \BibitemOpen
  \bibfield  {author} {\bibinfo {author} {\bibfnamefont {S.}~\bibnamefont
  {Muhammad}}, \bibinfo {author} {\bibfnamefont {A.}~\bibnamefont {Tavakoli}},
  \bibinfo {author} {\bibfnamefont {M.}~\bibnamefont {Kurant}}, \bibinfo
  {author} {\bibfnamefont {M.}~\bibnamefont {Paw\l{}owski}}, \bibinfo {author}
  {\bibfnamefont {M.}~\bibnamefont {\ifmmode~\dot{Z}\else \.{Z}\fi{}ukowski}},\
  and\ \bibinfo {author} {\bibfnamefont {M.}~\bibnamefont {Bourennane}},\
  }\bibfield  {title} {\bibinfo {title} {Quantum bidding in bridge},\ }\href
  {https://doi.org/10.1103/PhysRevX.4.021047} {\bibfield  {journal} {\bibinfo
  {journal} {Phys. Rev. X}\ }\textbf {\bibinfo {volume} {4}},\ \bibinfo {pages}
  {021047} (\bibinfo {year} {2014})}\BibitemShut {NoStop}%
\bibitem [{\citenamefont {Wilde}(2017)}]{Wilde2017}%
  \BibitemOpen
  \bibfield  {author} {\bibinfo {author} {\bibfnamefont {M.~M.}\ \bibnamefont
  {Wilde}},\ }\href@noop {} {\emph {\bibinfo {title} {Quantum Information
  Theory}}},\ \bibinfo {edition} {2nd}\ ed.\ (\bibinfo  {publisher} {Cambridge
  University Press},\ \bibinfo {year} {2017})\BibitemShut {NoStop}%
\bibitem [{\citenamefont {Barrett}\ and\ \citenamefont
  {Pironio}(2005)}]{Barrett2005}%
  \BibitemOpen
  \bibfield  {author} {\bibinfo {author} {\bibfnamefont {J.}~\bibnamefont
  {Barrett}}\ and\ \bibinfo {author} {\bibfnamefont {S.}~\bibnamefont
  {Pironio}},\ }\bibfield  {title} {\bibinfo {title} {Popescu-rohrlich
  correlations as a unit of nonlocality},\ }\bibfield  {journal} {\bibinfo
  {journal} {Physical Review Letters}\ }\textbf {\bibinfo {volume} {95}},\
  \href {https://doi.org/10.1103/physrevlett.95.140401}
  {10.1103/physrevlett.95.140401} (\bibinfo {year} {2005})\BibitemShut
  {NoStop}%
\bibitem [{\citenamefont {Tavakoli}\ \emph {et~al.}(2016)\citenamefont
  {Tavakoli}, \citenamefont {Marques}, \citenamefont {Paw\l{}owski},\ and\
  \citenamefont {Bourennane}}]{Tavakoli2016}%
  \BibitemOpen
  \bibfield  {author} {\bibinfo {author} {\bibfnamefont {A.}~\bibnamefont
  {Tavakoli}}, \bibinfo {author} {\bibfnamefont {B.}~\bibnamefont {Marques}},
  \bibinfo {author} {\bibfnamefont {M.}~\bibnamefont {Paw\l{}owski}},\ and\
  \bibinfo {author} {\bibfnamefont {M.}~\bibnamefont {Bourennane}},\ }\bibfield
   {title} {\bibinfo {title} {Spatial versus sequential correlations for random
  access coding},\ }\href {https://doi.org/10.1103/PhysRevA.93.032336}
  {\bibfield  {journal} {\bibinfo  {journal} {Phys. Rev. A}\ }\textbf {\bibinfo
  {volume} {93}},\ \bibinfo {pages} {032336} (\bibinfo {year}
  {2016})}\BibitemShut {NoStop}%
\bibitem [{\citenamefont {Brassard}\ \emph {et~al.}(2006)\citenamefont
  {Brassard}, \citenamefont {Buhrman}, \citenamefont {Linden}, \citenamefont
  {M\'ethot}, \citenamefont {Tapp},\ and\ \citenamefont
  {Unger}}]{Brassard2006}%
  \BibitemOpen
  \bibfield  {author} {\bibinfo {author} {\bibfnamefont {G.}~\bibnamefont
  {Brassard}}, \bibinfo {author} {\bibfnamefont {H.}~\bibnamefont {Buhrman}},
  \bibinfo {author} {\bibfnamefont {N.}~\bibnamefont {Linden}}, \bibinfo
  {author} {\bibfnamefont {A.~A.}\ \bibnamefont {M\'ethot}}, \bibinfo {author}
  {\bibfnamefont {A.}~\bibnamefont {Tapp}},\ and\ \bibinfo {author}
  {\bibfnamefont {F.}~\bibnamefont {Unger}},\ }\bibfield  {title} {\bibinfo
  {title} {Limit on nonlocality in any world in which communication complexity
  is not trivial},\ }\href {https://doi.org/10.1103/PhysRevLett.96.250401}
  {\bibfield  {journal} {\bibinfo  {journal} {Phys. Rev. Lett.}\ }\textbf
  {\bibinfo {volume} {96}},\ \bibinfo {pages} {250401} (\bibinfo {year}
  {2006})}\BibitemShut {NoStop}%
\bibitem [{\citenamefont {Tavakoli}\ and\ \citenamefont {\ifmmode~\dot{Z}\else
  \.{Z}\fi{}ukowski}(2017)}]{Zukowski2017}%
  \BibitemOpen
  \bibfield  {author} {\bibinfo {author} {\bibfnamefont {A.}~\bibnamefont
  {Tavakoli}}\ and\ \bibinfo {author} {\bibfnamefont {M.}~\bibnamefont
  {\ifmmode~\dot{Z}\else \.{Z}\fi{}ukowski}},\ }\bibfield  {title} {\bibinfo
  {title} {Higher-dimensional communication complexity problems: Classical
  protocols versus quantum ones based on bell's theorem or
  prepare-transmit-measure schemes},\ }\href
  {https://doi.org/10.1103/PhysRevA.95.042305} {\bibfield  {journal} {\bibinfo
  {journal} {Phys. Rev. A}\ }\textbf {\bibinfo {volume} {95}},\ \bibinfo
  {pages} {042305} (\bibinfo {year} {2017})}\BibitemShut {NoStop}%
\bibitem [{\citenamefont {Farkas}\ \emph {et~al.}(2025)\citenamefont {Farkas},
  \citenamefont {Miklin},\ and\ \citenamefont {Tavakoli}}]{Farkas2025}%
  \BibitemOpen
  \bibfield  {author} {\bibinfo {author} {\bibfnamefont {M.}~\bibnamefont
  {Farkas}}, \bibinfo {author} {\bibfnamefont {N.}~\bibnamefont {Miklin}},\
  and\ \bibinfo {author} {\bibfnamefont {A.}~\bibnamefont {Tavakoli}},\
  }\bibfield  {title} {\bibinfo {title} {Simple and general bounds on quantum
  random access codes},\ }\href {https://doi.org/10.22331/q-2025-02-25-1643}
  {\bibfield  {journal} {\bibinfo  {journal} {Quantum}\ }\textbf {\bibinfo
  {volume} {9}},\ \bibinfo {pages} {1643} (\bibinfo {year} {2025})}\BibitemShut
  {NoStop}%
\bibitem [{cod(2025)}]{code}%
  \BibitemOpen
  \href
  {https://github.com/esvegborn/Quantum_inputs_in_the_prepare_and_measure_scenario_and_stochastic_teleportation}
  {\bibinfo {title} {Code for optimal entangled assisted random access code}}
  (\bibinfo {year} {2025})\BibitemShut {NoStop}%
\bibitem [{\citenamefont {Tavakoli}\ \emph
  {et~al.}(2020{\natexlab{b}})\citenamefont {Tavakoli}, \citenamefont
  {Zambrini~Cruzeiro}, \citenamefont {Bohr~Brask}, \citenamefont {Gisin},\ and\
  \citenamefont {Brunner}}]{info1}%
  \BibitemOpen
  \bibfield  {author} {\bibinfo {author} {\bibfnamefont {A.}~\bibnamefont
  {Tavakoli}}, \bibinfo {author} {\bibfnamefont {E.}~\bibnamefont
  {Zambrini~Cruzeiro}}, \bibinfo {author} {\bibfnamefont {J.}~\bibnamefont
  {Bohr~Brask}}, \bibinfo {author} {\bibfnamefont {N.}~\bibnamefont {Gisin}},\
  and\ \bibinfo {author} {\bibfnamefont {N.}~\bibnamefont {Brunner}},\
  }\bibfield  {title} {\bibinfo {title} {Informationally restricted quantum
  correlations},\ }\href {https://doi.org/10.22331/q-2020-09-24-332} {\bibfield
   {journal} {\bibinfo  {journal} {{Quantum}}\ }\textbf {\bibinfo {volume}
  {4}},\ \bibinfo {pages} {332} (\bibinfo {year}
  {2020}{\natexlab{b}})}\BibitemShut {NoStop}%
\bibitem [{\citenamefont {Chaturvedi}\ and\ \citenamefont
  {Saha}(2020)}]{Chaturvedi2020}%
  \BibitemOpen
  \bibfield  {author} {\bibinfo {author} {\bibfnamefont {A.}~\bibnamefont
  {Chaturvedi}}\ and\ \bibinfo {author} {\bibfnamefont {D.}~\bibnamefont
  {Saha}},\ }\bibfield  {title} {\bibinfo {title} {Quantum prescriptions are
  more ontologically distinct than they are operationally distinguishable},\
  }\href {https://doi.org/10.22331/q-2020-10-21-345} {\bibfield  {journal}
  {\bibinfo  {journal} {{Quantum}}\ }\textbf {\bibinfo {volume} {4}},\ \bibinfo
  {pages} {345} (\bibinfo {year} {2020})}\BibitemShut {NoStop}%
\bibitem [{\citenamefont {Tavakoli}\ \emph {et~al.}(2022)\citenamefont
  {Tavakoli}, \citenamefont {Zambrini~Cruzeiro}, \citenamefont {Woodhead},\
  and\ \citenamefont {Pironio}}]{info2}%
  \BibitemOpen
  \bibfield  {author} {\bibinfo {author} {\bibfnamefont {A.}~\bibnamefont
  {Tavakoli}}, \bibinfo {author} {\bibfnamefont {E.}~\bibnamefont
  {Zambrini~Cruzeiro}}, \bibinfo {author} {\bibfnamefont {E.}~\bibnamefont
  {Woodhead}},\ and\ \bibinfo {author} {\bibfnamefont {S.}~\bibnamefont
  {Pironio}},\ }\bibfield  {title} {\bibinfo {title} {Informationally
  restricted correlations: a general framework for classical and quantum
  systems},\ }\href {https://doi.org/10.22331/q-2022-01-05-620} {\bibfield
  {journal} {\bibinfo  {journal} {{Quantum}}\ }\textbf {\bibinfo {volume}
  {6}},\ \bibinfo {pages} {620} (\bibinfo {year} {2022})}\BibitemShut {NoStop}%
\bibitem [{\citenamefont {Tavakoli}\ \emph {et~al.}(2019)\citenamefont
  {Tavakoli}, \citenamefont {Rosset},\ and\ \citenamefont
  {Renou}}]{Rosset2019}%
  \BibitemOpen
  \bibfield  {author} {\bibinfo {author} {\bibfnamefont {A.}~\bibnamefont
  {Tavakoli}}, \bibinfo {author} {\bibfnamefont {D.}~\bibnamefont {Rosset}},\
  and\ \bibinfo {author} {\bibfnamefont {M.-O.}\ \bibnamefont {Renou}},\
  }\bibfield  {title} {\bibinfo {title} {Enabling computation of correlation
  bounds for finite-dimensional quantum systems via symmetrization},\ }\href
  {https://doi.org/10.1103/PhysRevLett.122.070501} {\bibfield  {journal}
  {\bibinfo  {journal} {Phys. Rev. Lett.}\ }\textbf {\bibinfo {volume} {122}},\
  \bibinfo {pages} {070501} (\bibinfo {year} {2019})}\BibitemShut {NoStop}%
\bibitem [{\citenamefont {Ioannou}\ and\ \citenamefont
  {Rosset}(2022)}]{Ioannou2022}%
  \BibitemOpen
  \bibfield  {author} {\bibinfo {author} {\bibfnamefont {M.}~\bibnamefont
  {Ioannou}}\ and\ \bibinfo {author} {\bibfnamefont {D.}~\bibnamefont
  {Rosset}},\ }\bibfield  {title} {\bibinfo {title} {Noncommutative polynomial
  optimization under symmetry},\ }\Eprint {https://arxiv.org/abs/2112.10803}
  {arXiv:2112.10803 [quant-ph]}  (\bibinfo {year} {2022})\BibitemShut {NoStop}%
\bibitem [{\citenamefont {Bužek}\ and\ \citenamefont
  {Hillery}(1996)}]{Buzek1996}%
  \BibitemOpen
  \bibfield  {author} {\bibinfo {author} {\bibfnamefont {V.}~\bibnamefont
  {Bužek}}\ and\ \bibinfo {author} {\bibfnamefont {M.}~\bibnamefont
  {Hillery}},\ }\bibfield  {title} {\bibinfo {title} {Quantum copying: Beyond
  the no-cloning theorem},\ }\href {https://doi.org/10.1103/physreva.54.1844}
  {\bibfield  {journal} {\bibinfo  {journal} {Physical Review A}\ }\textbf
  {\bibinfo {volume} {54}},\ \bibinfo {pages} {1844–1852} (\bibinfo {year}
  {1996})}\BibitemShut {NoStop}%
\bibitem [{\citenamefont {Guo}\ \emph {et~al.}(2025)\citenamefont {Guo},
  \citenamefont {Tang}, \citenamefont {Pauwels}, \citenamefont {Cruzeiro},
  \citenamefont {Hu}, \citenamefont {Liu}, \citenamefont {Huang}, \citenamefont
  {Li}, \citenamefont {Guo},\ and\ \citenamefont {Tavakoli}}]{Guo2025}%
  \BibitemOpen
  \bibfield  {author} {\bibinfo {author} {\bibfnamefont {Y.}~\bibnamefont
  {Guo}}, \bibinfo {author} {\bibfnamefont {H.}~\bibnamefont {Tang}}, \bibinfo
  {author} {\bibfnamefont {J.}~\bibnamefont {Pauwels}}, \bibinfo {author}
  {\bibfnamefont {E.~Z.}\ \bibnamefont {Cruzeiro}}, \bibinfo {author}
  {\bibfnamefont {X.-M.}\ \bibnamefont {Hu}}, \bibinfo {author} {\bibfnamefont
  {B.-H.}\ \bibnamefont {Liu}}, \bibinfo {author} {\bibfnamefont {Y.-F.}\
  \bibnamefont {Huang}}, \bibinfo {author} {\bibfnamefont {C.-F.}\ \bibnamefont
  {Li}}, \bibinfo {author} {\bibfnamefont {G.-C.}\ \bibnamefont {Guo}},\ and\
  \bibinfo {author} {\bibfnamefont {A.}~\bibnamefont {Tavakoli}},\ }\bibfield
  {title} {\bibinfo {title} {Compression of entanglement improves quantum
  communication},\ }\bibfield  {journal} {\bibinfo  {journal} {Laser \&amp;
  Photonics Reviews}\ }\href {https://doi.org/10.1002/lpor.202401110}
  {10.1002/lpor.202401110} (\bibinfo {year} {2025})\BibitemShut {NoStop}%
\bibitem [{\citenamefont {Pimpel}\ \emph {et~al.}(2023)\citenamefont {Pimpel},
  \citenamefont {Renner},\ and\ \citenamefont {Tavakoli}}]{Pimpel2023}%
  \BibitemOpen
  \bibfield  {author} {\bibinfo {author} {\bibfnamefont {F.}~\bibnamefont
  {Pimpel}}, \bibinfo {author} {\bibfnamefont {M.~J.}\ \bibnamefont {Renner}},\
  and\ \bibinfo {author} {\bibfnamefont {A.}~\bibnamefont {Tavakoli}},\
  }\bibfield  {title} {\bibinfo {title} {Correspondence between entangled
  states and entangled bases under local transformations},\ }\href
  {https://doi.org/10.1103/PhysRevA.108.022220} {\bibfield  {journal} {\bibinfo
   {journal} {Phys. Rev. A}\ }\textbf {\bibinfo {volume} {108}},\ \bibinfo
  {pages} {022220} (\bibinfo {year} {2023})}\BibitemShut {NoStop}%
\bibitem [{\citenamefont {Pauwels}\ \emph {et~al.}(2024)\citenamefont
  {Pauwels}, \citenamefont {Pozas-Kerstjens}, \citenamefont {Santo},\ and\
  \citenamefont {Gisin}}]{Pauwels2025}%
  \BibitemOpen
  \bibfield  {author} {\bibinfo {author} {\bibfnamefont {J.}~\bibnamefont
  {Pauwels}}, \bibinfo {author} {\bibfnamefont {A.}~\bibnamefont
  {Pozas-Kerstjens}}, \bibinfo {author} {\bibfnamefont {F.~D.}\ \bibnamefont
  {Santo}},\ and\ \bibinfo {author} {\bibfnamefont {N.}~\bibnamefont {Gisin}},\
  }\bibfield  {title} {\bibinfo {title} {Classification of joint quantum
  measurements based on entanglement cost of localization},\ }\Eprint
  {https://arxiv.org/abs/2408.00831} {arXiv:2408.00831 [quant-ph]}  (\bibinfo
  {year} {2024})\BibitemShut {NoStop}%
\bibitem [{\citenamefont {Cavalcanti}\ \emph {et~al.}(2017)\citenamefont
  {Cavalcanti}, \citenamefont {Skrzypczyk},\ and\ \citenamefont {\ifmmode
  \check{S}\else \v{S}\fi{}upi\ifmmode~\acute{c}\else
  \'{c}\fi{}}}]{Cavalcanti2017}%
  \BibitemOpen
  \bibfield  {author} {\bibinfo {author} {\bibfnamefont {D.}~\bibnamefont
  {Cavalcanti}}, \bibinfo {author} {\bibfnamefont {P.}~\bibnamefont
  {Skrzypczyk}},\ and\ \bibinfo {author} {\bibfnamefont {I.}~\bibnamefont
  {\ifmmode \check{S}\else \v{S}\fi{}upi\ifmmode~\acute{c}\else \'{c}\fi{}}},\
  }\bibfield  {title} {\bibinfo {title} {All entangled states can demonstrate
  nonclassical teleportation},\ }\href
  {https://doi.org/10.1103/PhysRevLett.119.110501} {\bibfield  {journal}
  {\bibinfo  {journal} {Phys. Rev. Lett.}\ }\textbf {\bibinfo {volume} {119}},\
  \bibinfo {pages} {110501} (\bibinfo {year} {2017})}\BibitemShut {NoStop}%
\end{thebibliography}%

\clearpage
\onecolumngrid
\begin{appendix}
\pagenumbering{alph}

\section{Proof of Result \ref{Res:uni_fid} }\label{App:Worst_case_fid}
We give an explicit derivation of the universal fidelity in Result \ref{Res:uni_fid}. In the scenario of interest we stochastically want to teleport Alice's two unknown input qubits $ \psi = |\psi_1\rangle \otimes |\psi_2\rangle$ to Bob, using only 2 classical bits of communication and a two-qubit EPR pair $|\phi^+\rangle_{AB} = \frac{1}{\sqrt{2}}(|00\rangle + |11\rangle)$. To simplify the calculations we start by writing the total initial state of Alice's inputs and the shared entangled state $|\Psi\rangle= |\psi_1\rangle \otimes |\psi_2\rangle \otimes |\phi^+\rangle_{AB}$ in the Bell basis;
\begin{equation}
|\Psi \rangle = \sum_{i,j = 0,1} a_{ij} (\mathds{1} \otimes X^i Z^j)|\phi^+\rangle_{A'}  \otimes  |\phi^+\rangle_{AB},
\end{equation}
where the coefficients $a_{ij} \equiv \langle \psi_1 \psi_2 | \mathds{1} \otimes X^i Z^j|\phi^+\rangle_{A'}$ for all $i,j$  with $X$ and $Z$ the Pauli matrices. Alice then performs the multiparticle measurement $M^{c}_{A'A}$, with outcome $c = c_0c_1 \in \{0,1\}^2$, jointly on her two input particles and her share of the maximally entangled state. The measurements are defined as
\begin{equation}\label{eq:M_a}
M^c = \sum_{k= 0,1} |\psi_{c_0 c_1}^k\rangle \langle \psi_{c_0 c_1}^k| \,,
\end{equation}
where the states $ |\psi_{c_0 c_1}^k\rangle$ are given by
\begin{equation}\label{eq:psi_ak}
|\psi^k_{c_0 c_1}\rangle = X^{c_1+c_0+k}Z^{c_1} \otimes X^{c_1+k}Z^{c_0} \otimes X^k |\psi^{0}_{00}\rangle \,,
\end{equation}
and we the fiducial state $|\psi_{00}^0\rangle$ is given by
\begin{equation}
|\psi^{0}_{00}\rangle = \sqrt{\dfrac{2}{3}} |00\rangle_{A'} |1\rangle_{A} - \dfrac{1}{\sqrt{3}} |\psi^+\rangle_{A'} |0\rangle_{A} \,.
\end{equation}
The resulting subnormalised state, remotely prepared by Alice for Bob, takes the form
\begin{equation}
\sigma_{c\mid \psi} = \sum_{k= 0,1} \langle \psi_c^k \otimes \mathds{1}_B | \Psi\rangle  \langle \Psi | \psi_c^k \otimes \mathds{1}_B \rangle = \sum_{k=0,1} |\sigma_c^k\rangle \langle \sigma_c^k|,
\end{equation}
where each component $|\sigma^k_c\rangle = \langle \psi_c^k \otimes \mathds{1}_B | \Psi\rangle$ can be expressed as
\begin{equation}\label{eq:meas_states}
\begin{aligned}
|\sigma_c^k\rangle = \langle \psi_{00}^0 | \otimes \mathds{1}_B \bigg (\sum_{i,j =0,1} a_{ij} (-1)^{c_0(1+i+j) + j(c_1+k)} \big[(\mathds{1} \otimes X^{c_0+i}Z^{c_0+c_1+j})|\phi^+\rangle_{A'} \otimes (\mathds{1} \otimes X^k) |\phi^+\rangle_{AB}\big] \bigg).
\end{aligned}
\end{equation}

Explicitly, conditioned on the classical message $c$, the subnormalized states remotely prepared at Bob $\sigma_{c|\psi}$ are given by
\begin{equation}
\begin{aligned}
\sigma_{00\mid \psi} &= \dfrac{1}{6}
\begin{bmatrix}
|a_{10}|^2 + |a_{00}-a_{01}|^2 & -2\Re(a_{00}a_{10}^*)+ 2i\Im(a_{01}a_{10}^*)  \\
-2\Re(a_{00}a_{10}^*)- 2i \Im(a_{01}a_{10}^*) & |a_{10}|^2 + |a_{00}+a_{01}|^2
\end{bmatrix}\\
\\
\sigma_{01 \mid \psi } &= \dfrac{1}{6}
\begin{bmatrix}
|a_{11}|^2 + |a_{00}+a_{01}|^2 & -2\Re(a_{01}a_{11}^*)- 2i \Im(a_{00}a_{11}^*)  \\
-2\Re(a_{01}a_{11}^*)+ 2i \Im(a_{00}a_{11}^*) & |a_{11}|^2 + |a_{00}-a_{01}|^2
\end{bmatrix}\\
\\
\sigma_{10 \mid \psi} &= \dfrac{1}{6}
\begin{bmatrix}
|a_{01}|^2 + |a_{10}+a_{11}|^2 & 2\Re(a_{01}a_{11}^*) - 2i \Im(a_{01}a_{10}^*)  \\
 2\Re(a_{01}a_{11}^*) + 2i \Im(a_{01}a_{10}^*)  & |a_{01}|^2 + |a_{10}-a_{11}|^2
\end{bmatrix}\\
\\
\sigma_{11 \mid \psi} &= \dfrac{1}{6}
\begin{bmatrix}
|a_{00}|^2 + |a_{10}-a_{11}|^2 & 2\Re(a_{00}a_{10}^*) + 2i \Im(a_{00}a_{11}^*)  \\
 2\Re(a_{00}a_{10}^*) - 2i \Im(a_{00}a_{11}^*) & |a_{00}|^2 + |a_{10}+a_{11}|^2
\end{bmatrix}
\end{aligned}
\end{equation}

Alice sends the outcome of her measurement $c = c_0c_1$ to Bob as a classical message. Based on this message and his choice $y \in [2]$ Bob perform the unitary operation
\begin{equation}
U^{c,y} =  X^{1+yc_0+c_1}Z^{1+  (1+y)c_0+yc_1} \,.
\end{equation}
Bob's output state averaged over Alice's message $c$, then reads
\begin{equation}
\begin{aligned}
\tau_{y, \psi} &= XZ\sigma_{00 \mid \psi}(XZ)^\dagger + Z^{1+y} \sigma_{01 \mid \psi}( Z^{1+y})^\dagger+X^{1+y}Z^y \sigma_{10 \mid \psi} (X^{1+y}Z^y)^\dagger + X^y\sigma_{11 \mid \psi} ( X^y)^\dagger.
 \end{aligned}
\end{equation}
Let's focus on $y=1$ (the $y=2$ case works analogously). To compare $\tau_{1, \psi}$ to the input state  $\psi_1$, we first perform a change of basis. Write the unknown input qubits on the form $|\psi_1\rangle = a|0\rangle + b |1\rangle$ and $|\psi_2\rangle = c |0\rangle + d|1\rangle$, where $|a|^2+|b|^2 = 1, |c|^2+|d|^2 = 1$. These are related to the coefficients in the Bell basis $\{a_{ij}\}_{i,j=0,1}$ by
\begin{equation}
a_{00} = \dfrac{ac+bd}{\sqrt{2}}, \quad a_{01} = \dfrac{ac-bd}{\sqrt{2}}, \quad a_{10} = \dfrac{ad+bc}{\sqrt{2}}, \quad a_{11} = \dfrac{ad-bc}{\sqrt{2}}
\end{equation}
Inserting these coefficients into the expressions for $\sigma_{c \mid \psi}$, and the computing the expression $\tau_1= XZ\sigma_{00 \mid \psi}(XZ)^\dagger + \sigma_{01\mid \psi}+Z \sigma_{10 \mid \psi} Z^\dagger + X\sigma_{11\mid \psi} X^\dagger$, we find that
\begin{equation}
\begin{aligned}
\tau_{1,\psi} = \dfrac{1}{3}
\begin{bmatrix}
\dfrac{5}{2}|a|^2+\dfrac{1}{2}|b|^2 & 2a^*b  \\
2ab^* & \dfrac{5}{2}|b|^2+\dfrac{1}{2}|a|^2
\end{bmatrix}
 = \dfrac{5}{6} 
\begin{bmatrix}|a|^2 & a^*b  \\
ab^* & |b|^2
\end{bmatrix}
+\dfrac{1}{6} 
\begin{bmatrix}|b|^2 & -ab^*  \\
-a^*b & |a|^2
\end{bmatrix}= \dfrac{5}{6} \psi_1 + \dfrac{1}{6} \psi_1^{\perp}
\end{aligned}
\end{equation}
where $\psi_{1}^{\perp}$ is a state orthonormal to $\psi_{1}$. Similarly, by selecting $y = 2$ we find that $\tau_{2,\psi} = \dfrac{5}{6}\psi_{2} + \dfrac{1}{6}\psi_{2}^{\perp}$. Hence, the output state take the simple form
\begin{equation}
\tau_{y,\psi} = \sum_c U^{c,y} (\sum_k |\sigma^k_{c} \rangle \langle \sigma^k_{c}|) {U^{c,y}}^\dagger = \dfrac{5}{6} \psi_{y} + \dfrac{1}{6}\psi_{y}^\perp.
\end{equation}

\end{appendix}

\end{document}